\documentclass[aps,prx,floatfix,twocolumn,superscriptaddress,longbibliography]{revtex4-1}
\usepackage[english]{babel} 
\usepackage[latin1]{inputenc}
\usepackage[usenames,dvipsnames]{color}

\usepackage{amsmath,amssymb,amsfonts}
\usepackage[colorlinks=true,linkcolor=blue,urlcolor=blue,citecolor=blue]{hyperref}
\usepackage[percent]{overpic}
\usepackage[latin1]{inputenc}
\usepackage{amsmath,amssymb}
\usepackage{graphicx,color}
\usepackage{braket}
\usepackage{bm}
\usepackage[caption = false]{subfig}

\usepackage{titlesec}
\titleformat{\section}[runin]{\normalfont\itshape}{}{3pt}{}[.]

\newcommand{\hc}[0]{\textrm{H.c.}}

\newcommand{\trm}[1]{\textrm{#1}}

\begin{document}
\title{Magnetically-Confined Bound States in Rashba Systems}

\author{Flavio Ronetti}
%\email{flavio.ronetti@unibas.ch}
\affiliation{Department of Physics, University of Basel, Klingelbergstrasse 82, CH-4056 Basel, Switzerland}
\author{Kirill Plekhanov}
\affiliation{Department of Physics, University of Basel, Klingelbergstrasse 82, CH-4056 Basel, Switzerland}
\author{Daniel Loss}
\affiliation{Department of Physics, University of Basel, Klingelbergstrasse 82, CH-4056 Basel, Switzerland}
\author{Jelena Klinovaja}
\affiliation{Department of Physics, University of Basel, Klingelbergstrasse 82, CH-4056 Basel, Switzerland}

\begin{abstract} A Rashba nanowire is subjected to a magnetic field that assumes opposite signs in two sections of the nanowire, and, thus, creates a magnetic domain wall. The direction of magnetic field is chosen to be perpendicular to  the Rashba spin-orbit vector such that there is only a partial gap in the spectrum. Nevertheless, we prove analytically and numerically that such a domain wall hosts a bound state whose energy is at bottom of the spectrum below the energy of all bulk states. Thus, this magnetically-confined bound state is well-isolated and can be accessed experimentally. We further show that the same type of magnetic confinement  can be implemented in two-dimensional systems with strong spin-orbit interaction. A quantum channel  along the magnetic domain wall emerges
%A one-dimensional nondegenerate quantum channel, corresponding to states 
with a non-degenerate dispersive band that lies energetically below the bulk states.
%is localized at the domain wall. 
We show that this magnetic confinement is robust against disorder and various parameter variations.
\end{abstract}

\maketitle

\textit{Introduction.} The possibility of confining electrons to manipulate their quantum state plays an extremely important role in condensed matter physics and paves the way for various quantum computing schemes \cite{Loss98,Kitaev2001,Fowler09,Kloeffel13}. The prime example are quantum dots where the confinement can be generated by external gates or intrinsically via mismatch of band gaps. The confinement can also result from non-uniform superconducting gaps giving rise to Andreev bound states~\cite{Sauls18,Lee12,Grove18,Junger19,Hays18}. Other ways to confine states are based on interfaces or domain walls which separate regions of different phases, well-known examples being Jackiw-Rebbi fermions \cite{Jackiw1976,Su1979,Rajaraman1982,Kivelson1982,Heeger1988,Klinovaja2013,Deng14,Rainis2014,Nishida10} and, in particular, Majorana bound states in proximitized nanowires with Rashba spin-orbit interaction (SOI)~\cite{Oreg2010,Lutchyn2010,Potter2011,Sticlet2012,Halperin2012,San-Jose2012,Rainis2013,Das2012,Deng2012,Lutchyn18,Deng16,Deng18,Mourik12}. 

It is then natural to ask if there are further ways to confine electrons and thereby open up new platforms for bound states.
Motivated by this question, we consider  systems with uniform Rashba SOI  in the presence of a non-uniform magnetic field with a domain wall.
In case of a nanowire (NW) where the direction of the magnetic field is perpendicular to the  Rashba SOI  vector, a partial gap opens in the spectrum~\cite{Streda2003,Pershin2004,Meng2013b,Kammhuber2017}. 
Naively linearizing the interior branches around zero momentum \cite{Klinovaja2015}, one might expect that these branches can be mapped to the Jackiw-Rebbi model that would result in a bound state  in the middle of the partial gap coexisting with the extended states from the outer (ungapped) branches.
However, despite the fact that there is a gap inversion at the domain wall, we do not find any localized states in this approach. Quite surprisingly, if we go beyond linearization and take band curvature effects into account, a bound state does emerge  that lies now not inside the partial gap but at the bottom of the spectrum, below all extended states.  
Remarkably, such a magnetically-confined bound state occurs even in the regime where the Zeeman energy is much smaller than the SOI energy. While for analytical calculations a sharp transition of magnetic field is considered, numerically we confirm that  the bound state and bulk states are energetically well-separated even for smooth magnetic domain walls. We also show that the bound states are robust against disorder and various parameter variations. Finally, we consider a two-dimensional (2D) Rashba layer and show that, similarly to the NW, a  one-dimensional quantum channel, whose dispersion lies energetically below any other bulk states, arises at the interface between two regions of opposite perpendicular magnetic fields. The breaking of the inversion symmetry in the spectrum opens access to the ratio between Rashba and Dresselhaus SOI terms.
The  setups proposed here can be experimentally implemented by placing a Rashba system on ferromagnets with magnetic domains.

\begin{figure}[t]
	\includegraphics[width = \linewidth]{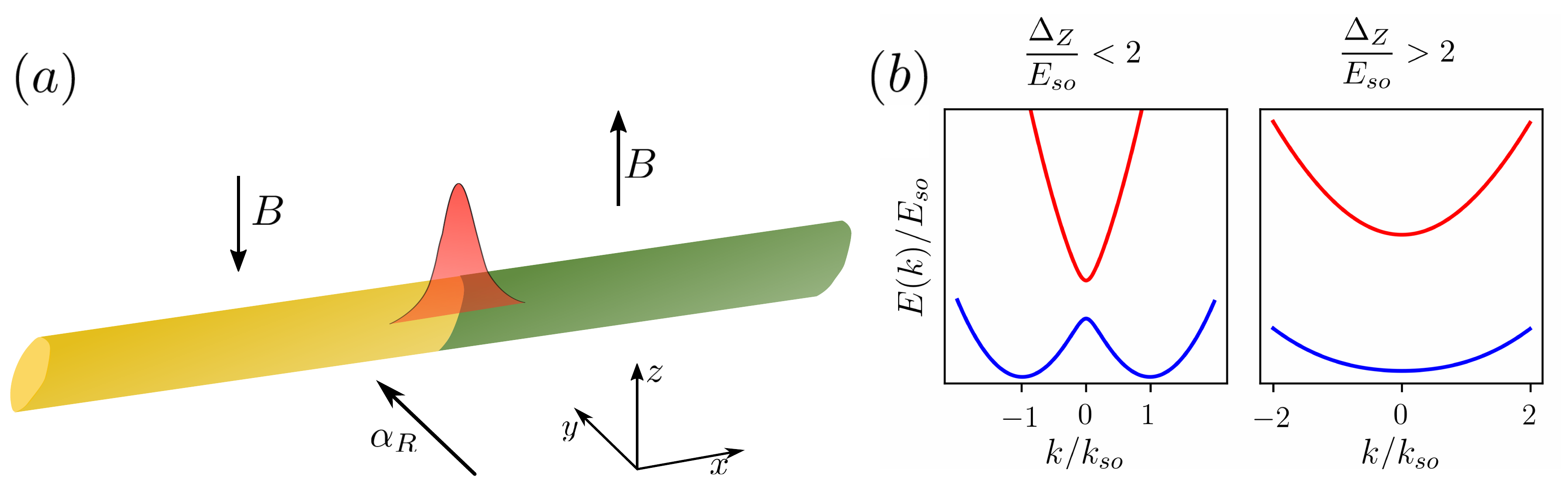}
	\caption{ (a) The NW aligned in $x$ direction with uniform Rashba SOI vector $\boldsymbol{\alpha}_R$  pointing along   $y$-axis. A magnetic field $\mathbf{B}$ is applied perpendicular to $\boldsymbol{\alpha}_R$, {\it i.e. } along $z$-axis, resulting in a partial gap in the bulk spectrum. To generate a  magnetic domain wall hosting a bound state (red), the directions of $\mathbf{B}$ are chosen to be opposite for $x>0$ (green region) and for $x<0$ (yellow region). (b)  Bulk spectrum for uniform magnetic field for two different regimes of relative strength between Zeeman energy $\Delta_Z$ and Rashba SOI energy $E_{so}$. A gap at zero momentum $k=0$ given by $\Delta_Z$ separates the two bands.  If $\Delta_Z/E_{so}<2$,  the lowest band has a local maximum at $k=0$ and  a local minimum around $k\sim k_{so}$. If $\Delta_Z/E_{so}>2$, only a single global minimum exists at $k=0$.}
	\label{fig:setup}
\end{figure}

\textit{Model.} We consider a Rashba  NW aligned along the $x$-axis and subjected to a non-uniform magnetic field, see Fig. \ref{fig:setup}.
The kinetic part of the Hamiltonian is given by (with $\hbar=1$)
\begin{equation}
H_0=
%\sum_{\sigma,\sigma'}
\int dx\ \Psi^{\dagger}_{\sigma}(x)\left[\frac{- \partial_x^2}{2m}-\mu-i\alpha_R\sigma_y \right]_{\sigma\sigma'}\Psi_{\sigma'}(x),
\end{equation}
where $\Psi_{\sigma}(x)$ is the annihilation operator acting on an electron with spin $\sigma/2=\pm 1/2$ at position $x$ of the NW and $\sigma_i$ is the Pauli matrix acting on the electron spin. Here, $\mu$ is the chemical potential and $m$  the effective mass. The Rashba SOI, assumed to be uniform, is characterized by the SOI vector $\boldsymbol{\alpha}_R$, which is  aligned along the $y$ direction. In addition, we also define the SOI momentum (energy) $k_{so}=m \alpha_R$  ($E_{so}=k_{so}^2/2m$).

To generate a domain wall, we apply an external magnetic field $\mathbf B$ perpendicular to the SOI vector $\boldsymbol{\alpha}_R$, i.e. along the $z$-axis. We assume that  $\mathbf B$ has opposite directions in the two regions $x>0$ and $x<0$, thus, allowing for the existence of localized bound states at the domain wall at $x=0$. In order to address this non-uniform magnetic field, we  introduce a position-dependent Zeeman term  given by
\begin{equation}
H_Z=
%\sum_{\sigma,\sigma'}
\int dx \hspace{1mm}\Delta_Z(x) \Psi^{\dagger}_{\sigma}(x)\left(\sigma_z\right)_{\sigma\sigma'}\Psi_{\sigma'}(x).
\end{equation}
To provide an analytical treatment of this model, we focus on a specific functional dependence of the Zeeman energy, $\Delta_Z(x)=\Delta_Z {\text{sgn}(x)}$ with $\Delta_Z=g \mu_B B$, where $g$ is the $g$-factor and $\mu_B$  the Bohr magenton. This particular choice mimicks an abrupt change of direction at the interface $x=0$. The effects of a smooth transition  can  be treated by  numerical simulations, where the smooth change of the  magnetic field can be described e.g. as $\Delta_Z(x)=\Delta_Z \tanh(x/\delta)$. Here, the parameter $\delta$ characterizes the width of the domain wall. The abrupt change at $x=0$ corresponds to the limit  $ k_{so}\delta\rightarrow0$.

We note that the configuration described above could be mapped to an equivalent system without Rashba SOI by applying the spin-dependent gauge transformation $\Psi_{\sigma}(x)\rightarrow e^{i\sigma k_{so}x}\Psi_{\sigma}(x)$~\cite{Braunecker2010}. This transformation eliminates the term $H_R$ in the Hamiltonian and changes the Zeeman energy as $\Delta_Z(x)\rightarrow \Delta_Z(x)\left[\cos(2k_{so}x)\hat{z}+\sin(2k_{so}x)\hat{x}\right]$, which corresponds to a helical magnetic field, which could be created either extrinsically by arrays of nanomagnets \cite{Braunecker2010,Karmakar11,Klinovaja2012,Fatin16,Abiague17, Maurer18, Mohanta19,Desjardins2019} 
 or intrinsically via ordering of nuclear spins or magnetic adatoms due to RKKY interaction \cite{
 Braunecker09b,Scheller14,Hsu15}. By analogy, the domain walls occurring in such structures will also host bound states, see the Supplemental Material (SM) \cite{supp}.

The total Hamiltonian is given by $H=H_0+H_Z$, and 
%\equiv \int dx\ \mathcal{H}(x)$.
%where $\mathcal{H}(x)$ is the associated Hamiltonian density. 
for a uniform magnetic field $\Delta_Z(x)\equiv \Delta_Z$ its energy spectrum consists of two bands separated by a gap,
\begin{equation}
\label{eq:rashba}
E_{\pm}(k)=\left(k^2+k_{so}^2\pm2 \sqrt{k^2 k_{so}^2+m^2 \Delta _Z^2}\right)/2 m.
\end{equation}
We note that the shape of these bands changes significantly depending on whether the dominant energy scale is the SOI energy $E_{so}$ or the Zeeman energy $\Delta_Z$.  In the first case, the lowest band $E_{-}(k)$ has a local maximum at $k=0$ as well as a two minima close to $k \sim k_{so}$.
In the opposite case, only a single  global mininum exists at $k=0$ [see Fig. \ref{fig:setup}(b)]. The transition between these two regimes occurs at $\Delta_Z/E_{so}=2$. In general, the bottom of the lowest band $E_{-}(k)$, denoted as $E_1$,  moves according to the following expression: 
\begin{equation}
E_1=\begin{cases}
-\frac{\Delta_Z^2}{4E_{so}},\ \ \ \ \ \ \ \  & \Delta_Z<2E_{so}\\
E_{so}-\Delta_Z, &\Delta_Z\ge 2E_{so}
\end{cases}.
\label{eq:energybottom}
\end{equation}
This value corresponds to the minimal energy of bulk electrons in the NW. Surprisingly, as we will show in the following, a bound state localized at the domain wall at $x=0$ can exist even at energies \textit{below} $E_1$. Let us emphasize that this problem cannot be tackled by linearizing the spectrum close to the Fermi energy and has to be solved by taking into account the exact parabolic dispersion of the NW.

\textit{Bound state at the interface.} In order to demonstrate the existence of a bound state at the interface $x=0$ for energy below the bulk spectrum analytically, one has to solve the Schroedinger equation $\mathcal{H}(x)\psi(x)=E\psi(x)$, 
where we choose $E<0$ and focus on solutions below the bottom of the band $E<E_1$. 
Here,   $\mathcal{H}(x)$ is  the Hamiltonian density associated with  $H=H_0+H_Z$, and $\psi(x)=\left (
\psi_{\uparrow}(x), \psi_{\downarrow}(x)
\right)^T$ is a 2D spinor. 
In addition, we consider a sharp domain wall with $\Delta_Z(x)=\Delta_Z {\text{sgn} (x)}$. The full solution
can be constructed from the solutions in the two different regions $x>0$ and $x<0$
by matching the corresponding wavefunctions at the interface $x=0$. The eigenfunction of $\mathcal{H}$ in each of the two separate regions has the following form: $\psi(x)=\left(
v_{\uparrow}(k), v_{\downarrow}(k)
\right)^Te^{i kx}$, where $k$ is a complex number obtained by solving the equation $E_{\pm}(k)=E$ in the regime $E<E_1$, see Eq. \eqref{eq:rashba}. Indeed, an exponential decay required for having a bound state is encoded in the imaginary part of $k$: the latter should be positive (negative) for $x>0$ ($x<0$), in order to find a normalizable solution to the Schroedinger equation localized at  $x=0$.
We find that there exists a non-degenerate bound state with energy $E_{BS}<E_1$ under the condition $\Delta_Z/E_{so}<4$.
The expression for the energy $E_{BS}$ of this bound state is involved but still can be found analytically:
\begin{widetext}
\begin{equation}
 \frac{E_{BS}}{E_{so}}=-\frac{2^{\frac{2}{3}} \left[3 \left(\frac{\Delta_Z}{E_{so}}\right)^2+2\right]+\frac{2^{\frac{1}{3}}}{12} \left[27 \left(\frac{\Delta_Z}{E_{so}}\right)^4+72 \left(\frac{\Delta_Z}{E_{so}}\right)^2+3\sqrt{81 \left(\frac{\Delta_Z}{E_{so}}\right)^8+48 \left(\frac{\Delta_Z}{E_{so}}\right)^6}+32\right]^{\frac{2}{3}}}{3 \sqrt[3]{27 \left(\frac{\Delta_Z}{E_{so}}\right)^4+72 \left(\frac{\Delta_Z}{E_{so}}\right)^2+3 \sqrt{81 \left(\frac{\Delta_Z}{E_{so}}\right)^8+48 \left(\frac{\Delta_Z}{E_{so}}\right)^6}+32}}+\frac{2}{3}.\label{eq:gs}
 \end{equation}
 \end{widetext}
If $\Delta_Z/E_{so}\geq 4$, the bound state disappears by merging with the bulk spectrum. 

\begin{figure}[b]
	\includegraphics[width = \linewidth]{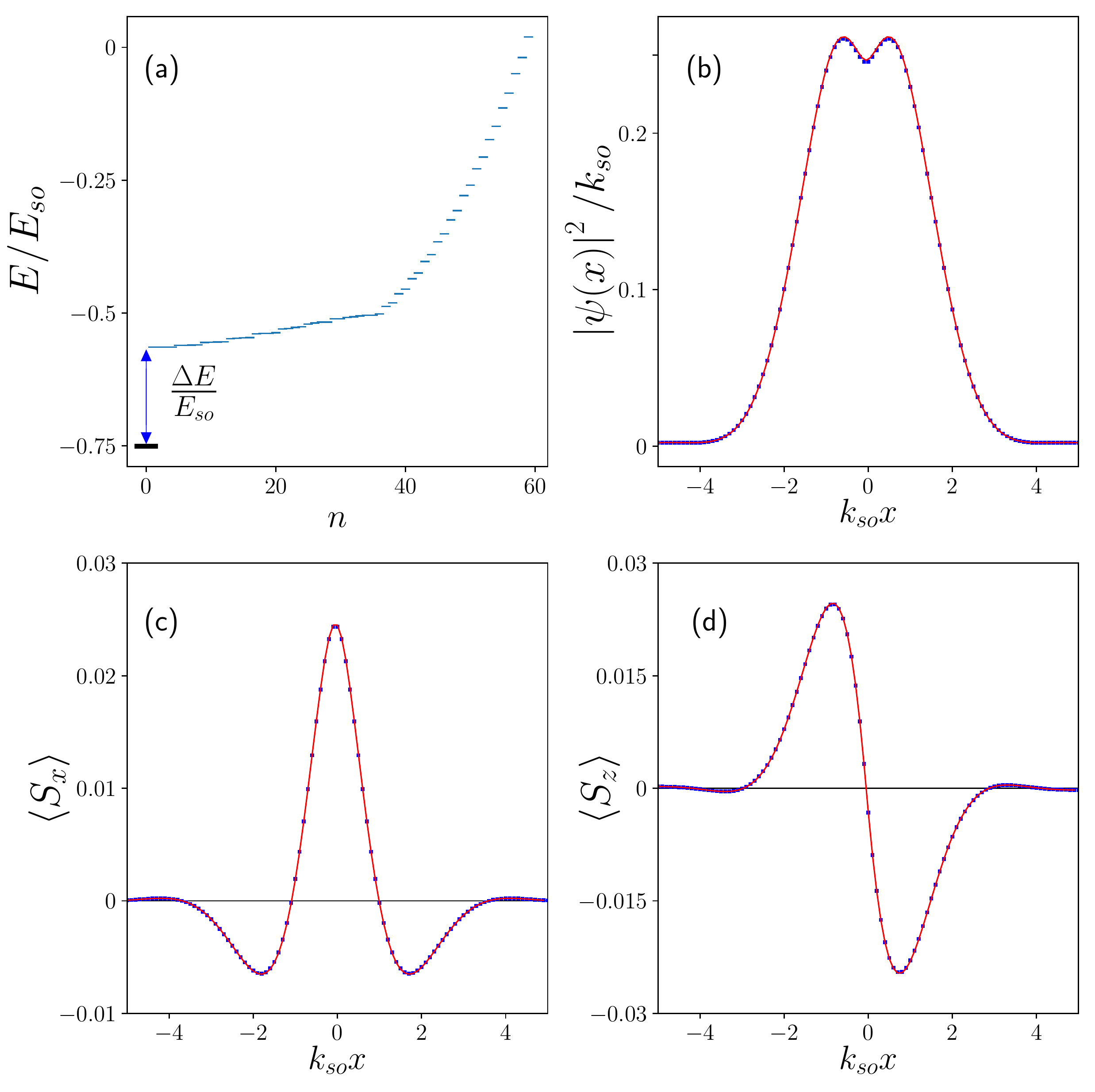}
	\caption{(a) The lowest energy  states of  a Rashba NW with a Zeeman field ($\Delta_Z/E_{so}=1.5$) that changes sign at $x=0$, obtained by numerical diagonalization of a system with $1300$ sites. The lowest energy state (black line), which corresponds to the non-degenerate bound state at $x=0$, is energetically well separated by $\Delta E$ from the bulk spectrum. (b) The probability density $\left|\psi(x)\right|^2$ of the bound state, (c) the $x$-component, and (d) the $z$-component of its spin polarization. The analytical  (red lines) and numerical results (blue dots) are in excellent agreement. }
	\label{fig:spectrum}
\end{figure}

\begin{figure}[h]
	\includegraphics[width = \linewidth]{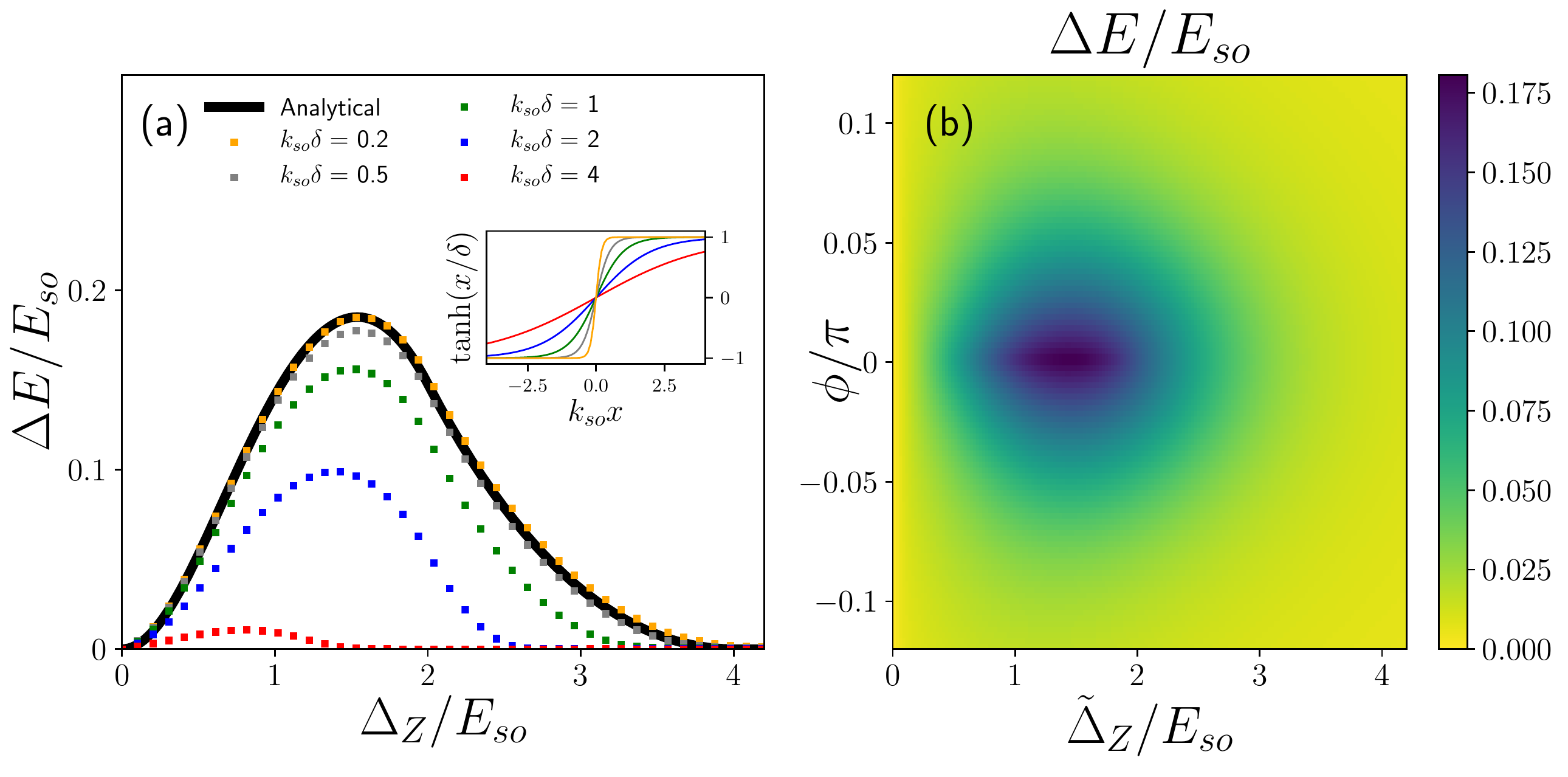}
	\caption{(a) Energy separation $\Delta E/E_{so}$ between the bound state and the lowest bulk state as a function of $\Delta_Z/E_{so}$ for different sizes of the  domain wall $\delta$. The different colors are assigned to points obtained by numerical simulation. Black solid line corresponds to the analytical result obtained in the sharp domain wall limit. (Inset) The magnetic field profile close to the interface for different value of $\delta$. The larger $\delta$, the smaller is the energy separation between the bound state and delocalized bulk states. Also, the smoother the profile, the smaller is $\Delta_Z$ at which the bound state merges with the continuum.
(b) Energy separation  $\Delta E/E_{so}$ as a function of the magnitude of magnetic field $\tilde{\Delta}_Z$ and the angle $\phi$ that the magnetic field forms with the $z$-axis in the presence of an additional parallel component. This shows the robustness of $\Delta E$ towards tilting of the magnetic field.}
	\label{fig:energies}
\end{figure}

Next, we define the energy separation $\Delta E=E_{1}-E_{BS}\geq0$ between this bound state and the lowest bulk state [see Fig. \ref{fig:spectrum}(a)].  If there is no bound state in the spectrum, we set $\Delta E=0$. In the limit of weak Zeeman field, the bound states splits from the bulk modes quadratically in the Zeeman energy as 
\begin{equation}
\Delta E=\frac{1}{4}\frac{\Delta_Z^2}{E_{so}}.
\end{equation}
Comparing the analytical solution and the numerical solution obtained in the discretized model, we find  excellent agreement [see Fig. \ref{fig:energies}].  The found bound state localized at $x=0$ is the lowest energy state  and is well-separated from then extended bulk modes.  We also confirm that, as expected, the bound state merges with the bulk states at $\Delta_Z/E_{so}=4$.

Even though we focused on the lowest two  bands of the NW,  identical bound states also appear right below the bottom of higher  band pairs. However, the visibility of such bound states is masked by the presence of extended bulk modes from lower bands~\cite{Hsu16,Kennes19}.\\ 

\textit{Bound state wavefunction and polarization.} Using numerical diagonalization, we can extract the  spectrum 
for arbitrary profiles of magnetic fields, and, moreover,  get access to the bound state wavefunction.
In case of a sharp domain wall, once the energy of the bound state has been obtained analytically, the analytical expression for the wavefunction of the  bound state can also be derived. The corresponding probability density is plotted in Fig. \ref{fig:spectrum}(b) and compared with the numerical solution. The agreement between the two quantities is excellent. From the analytics, we also obtain the localization length of the bound state
\begin{equation}
\xi_{BS}^{-1}=k_{so}\mathfrak{Im}\sqrt{1+\frac{E_{BS}}{E_{so}}+\sqrt{ \frac{4E_{BS}}{E_{so}}+\left(\frac{\Delta_Z} { E_{so}}\right)^2}},
\end{equation}
which, for the parameters of Fig. \ref{fig:spectrum}(b), is equal to $\xi_{BS}\sim 1.75 / k_{so}$.  Thus, we have established the 
existence of a bound state localized within $k_{so}^{-1}$ around $x=0$ and with an energy separation $\Delta E$ well below the energy of the bulk states. 
 
It is also interesting to study the  spin polarization of this bound state, $\left\langle S_{i}(x)\right\rangle=\sum_{\sigma,\sigma'}\psi_{\sigma}^*(x)\left(\sigma_{i}\right)_{\sigma\sigma'}\psi_{\sigma'}(x)$ with $i=x,y,z$.
We observe that the  polarization along the SOI vector $\boldsymbol{\alpha}_R$  vanishes, $\left\langle S_{y}(x)\right\rangle=0$, i.e., due to the mirror symmetry~\cite{Serina18},  the polarization stays orthogonal to $\boldsymbol{\alpha}_R$. The other two components are non-zero, see Fig. \ref{fig:spectrum}(c,d).
The $x$-component 
$\left\langle S_{x}(x)\right\rangle$ is symmetric with respect to $x=0$ with a global maximum close to the interface. Away from the interface, $\left\langle S_{x}(x)\right\rangle$ changes sign and reaches its global minimum before vanishing after a length of a few  $k_{so}^{-1}$. The component  along the magnetic field, $\left\langle S_{z }(x)\right\rangle$, is odd in $x$ and follows the sign of the magnetic field.

{\it Stability of the bound state.} Numerically, we can study the stability of the bound states away from the sharp domain wall limit. First, we  consider a  domain wall with smooth transition, modelled by $\Delta_Z(x)=\Delta_Z\tanh(x/\delta)$, see Fig. \ref{fig:energies}.
The agreement between analytical and numerical results improves as $\delta$ decreases and for $k_{so} \delta<0.2$ the match is almost exact. In the case of smoother transitions with $k_{so} \delta>0.2$, the bound state merges with the continuum  at smaller Zeeman energies (smaller than $\Delta_Z/E_{so}=4$). The analytical expression for the energy separation exhibits a maximum $\Delta E/E_{so}\sim 0.18 $ around $\Delta_Z/E_{so}\sim 1.5$. This value is reduced as $\delta$ grows. 
This opens the door to access the spin-orbit energy experimentally: measuring the value of $\Delta_Z$ at which the bound state disappears provides an estimate on $E_{so}$.

To confirm the robustness of the bound state, we also study numerically the case in which the magnetic field has a small uniform component $B_{||}$ 
parallel to the SOI vector.   As a result, the total field has an angle $\phi=\arctan(B_{||}/B)$ with the $z$-axis, see Fig. \ref{fig:energies}(b).
The corresponding Zeeman energy becomes $\tilde{\Delta}_Z=g \mu_B \sqrt{B^2+B_{||}^2}$, where, for simplicity, we assumed an isotropic $g$-factor. As  expected, the energy separation between bulk states and the bound state has a maximum  in the regime where the domain wall is most pronounced, $\phi=0$ (for $\tilde{\Delta}_Z/E_{so}\sim 1.5$). However, there  exists a wider range of  magnetic field orientations, which we estimate as  $|\phi|<\pi/10$, for which the bound states still exist.
In addition, we confirmed numerically the stability against disorder by allowing for fluctuations in chemical potential and magnetic field (see SM~\cite{supp}). All this demonstrates that the emergence of the bound states does note rely on fine-tuning of parameters but  is a rather stable effect.

\begin{figure}[t]
	\includegraphics[width = \linewidth]{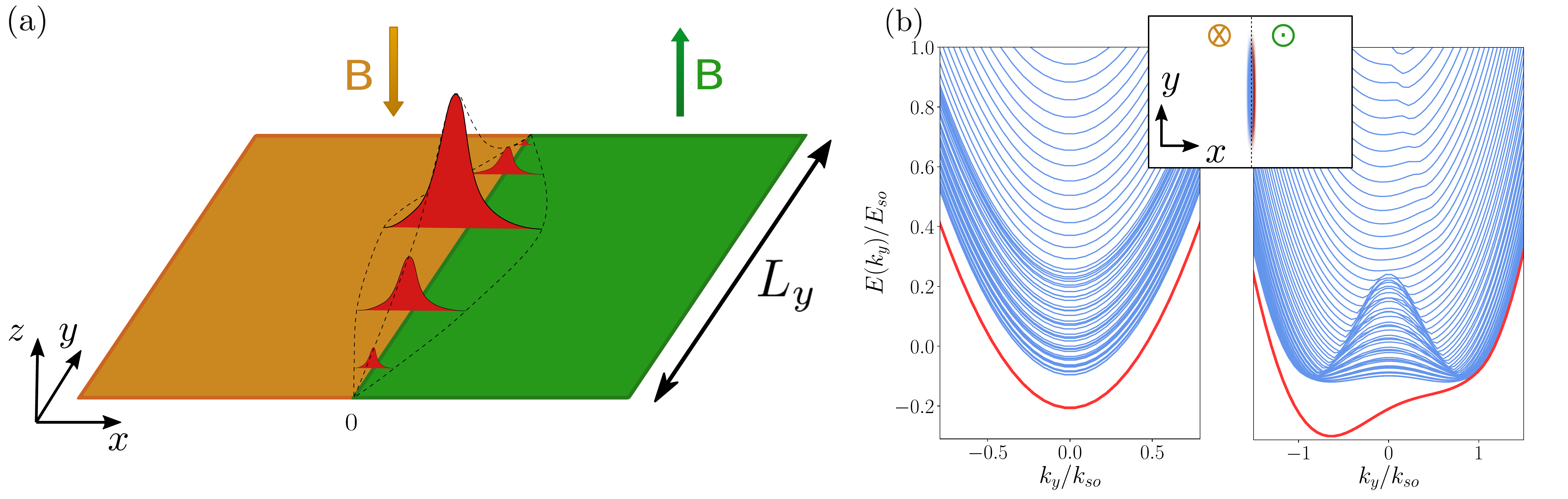}
\caption{(a) 2D Rashba layer placed in the $xy$-plane. 
A  Zeeman field $\mathbf{B}$ is applied perpendicular to the plane with opposite directions for  $x>0$ (green region) and for $x<0$ (yellow region) creating a one-dimensional domain wall at $x=0$. This domain wall hosts a quantum channel running  along the $y$-axis.  (b)  Energy spectrum as function of $k_y$ in the absence (presence) of Rashba SOI along the $y$-axis is shown on left (right) panel. The red curves describe the  spectrum of the non-degenerate quantum channel states localized along the domain wall. (Inset) Probability density of the quantum channel states found numerically for a finite-size system.
}
	\label{fig:2d}
\end{figure}

\textit{Quantum channel along domain wall in Rashba layer.} We extend now our consideration to 2D systems with strong SOI in the presence of a perpendicular magnetic Zeeman field whose sign is opposite in the two regions of the plane separated by the line $x=0$, defining the domain wall, see Fig. \ref{fig:2d}(a). We assume periodic boundary conditions in $y$ direction and, thus, the associated momentum $k_y$ is a good quantum number.  In two dimensions, a bound state localized along the domain wall  evolves into an extended one-dimensional channel  with the dispersive energy spectrum, see Fig. \ref{fig:2d}(b).

We consider two configurations for the 2D SOI described by $\mathcal{H}_R=-i \left(\alpha_x \sigma_y\partial_y-\alpha_y \sigma_x\partial_x\right)$.
In the first one, corresponding to the left panel of Fig. \ref{fig:2d}(b), the Rashba and Dresselhaus SOI are equal resulting in $\alpha_y=0$~\cite{Nitta97,Engels97,Schliemann03,Winkler03,Bernevig06,Meier07,Studer09,Koralek09,Meng14}. 
Here, the lower energy states, defining the 1D quantum channel, have a finite separation from the 2D bulk states (like the bound state in the NW). In this case, the energy dispersion of the channel states acquires a simple parabolic form, $E(k_y)=E_{BS}+{k_y^2}/{2m}$, which is symmetric with respect to $k_y=0$, with $E_{BS}$ given in Eq.~(\ref{eq:gs}), see the left panel of Fig. \ref{fig:2d}(b).
If both $\alpha_x$ and $\alpha_y$ are finite, the quantum channel dispersion relation is now asymmetric with respect to $k_y$,  see the right panel of Fig. \ref{fig:2d}(b). The largest energy separation occurs at a finite value of momentum and acquires a larger value compared to the previous case. Interestingly, this asymmetry in the energy dispersion  provides a test for the presence of a second component of SOI and a way to access its magnitude. 
Finally, the probability density in real space has the same  shape for both configurations of Rashba interaction [see Fig. \ref{fig:2d}(b)]:  as expected, the quantum channel is extended along the domain wall at $x=0$. We also verified that this state is still localized along the domain wall even for curved or closed boundaries of the wall, as long as there is no large in-plane magnetic field parallel to one of the SOI components (see SM~\cite{supp}).

\textit{Conclusions.}  We considered a Rashba NW in the presence of a domain wall created by a perpendicular magnetic Zeeman field, with opposite sign in the two corresponding domains. At the domain wall, a bound state exists whose energy is separated from the lowest bulk modes. This separation persists for smooth domain walls and for a slightly tilted magnetic field. This effect is straightforwardly extended to 2D Rashba layers where the domain wall hosts a quantum channel: a propagating non-degenerate mode with parabolic dispersion.
Our predictions can be tested by transport \cite{Grove18,Junger19,Hays18} and cavity measurements~\cite{Cubaynes19} in an experimental configuration with a spatially oscillating magnetic texture, which is equivalent to the presented setup. This texture could be produced with several different mechanisms. First, it could be obtained by making use of extrinsic nanomagnets~\cite{Braunecker2010,Karmakar11,Klinovaja2012,Fatin16,Abiague17, Maurer18, Mohanta19}. 
Second, one can implement a helical magnetic field in a system with local magnetic moments, such as nuclear spins or magnetic impurities~\cite{
Braunecker09b,Scheller14,Hsu15}. Finally, moving the magnetic domain walls adiabatically will allow to move the bound states attached to them.

\textit{Acknowledgments.} This work was supported by the Swiss National Science Foundation and NCCR QSIT. This project received funding from the European Union's Horizon 2020 research and innovation program (ERC Starting Grant, grant agreement No 757725).

\clearpage
\widetext
\begin{center}
	\textbf{\large Supplemental Material: Magnetically-Confined Bound States in Rashba Systems}\\
	\vspace{8pt}
	Flavio Ronetti,$^{1}$ Kirill Plekhanov,$^{1}$ 
	Daniel Loss,$^{1}$ and Jelena Klinovaja$^{1}$ \\ \vspace{4pt}
	$^{1}$ {\it Department of Physics, University of Basel,
		Klingelbergstrasse 82, CH-4056 Basel, Switzerland}
\end{center}

%% Merge with supplemental materials
%% Prefix a "S" to everything and reset the counter
\setcounter{section}{0}
\setcounter{equation}{0}
\setcounter{figure}{0}
\setcounter{page}{1}
\makeatletter
\renewcommand{\thesection}{S\arabic{section}}
\renewcommand{\theequation}{S\arabic{equation}}
\renewcommand{\thefigure}{S\arabic{figure}}
\titleformat{\section}[hang]{\large\bfseries}{\thesection.}{5pt}{}
\section{\label{secSm:boundstate}Bound state energy and wavefunction}

\begin{figure}[b]
	\subfloat[Imaginary part of $k_{+,+}/k_{so}$]{\includegraphics[width = 0.3\linewidth]{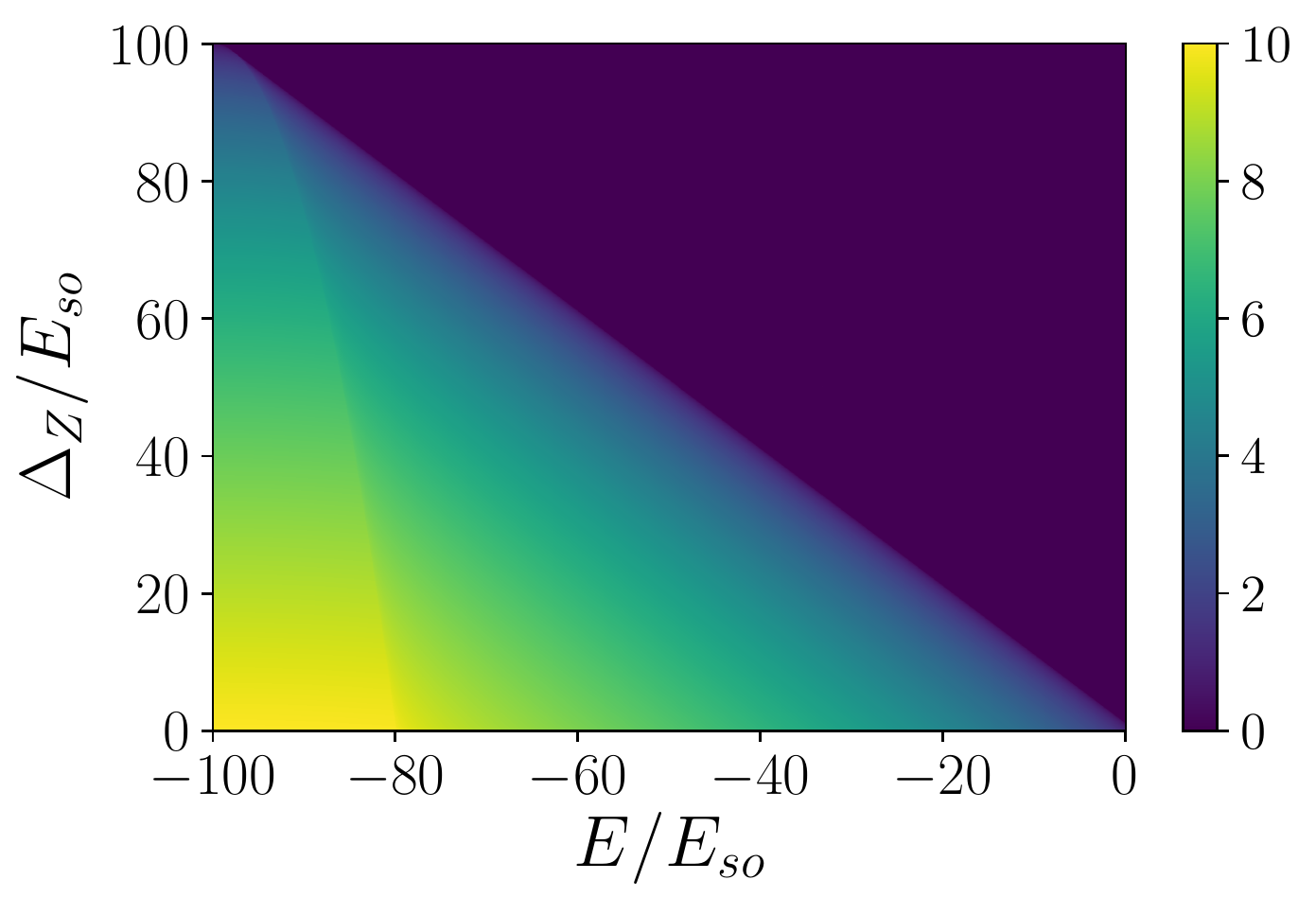}} 
	\subfloat[Imaginary part of $k_{-,-}/k_{so}$]{\includegraphics[width = 0.3\linewidth]{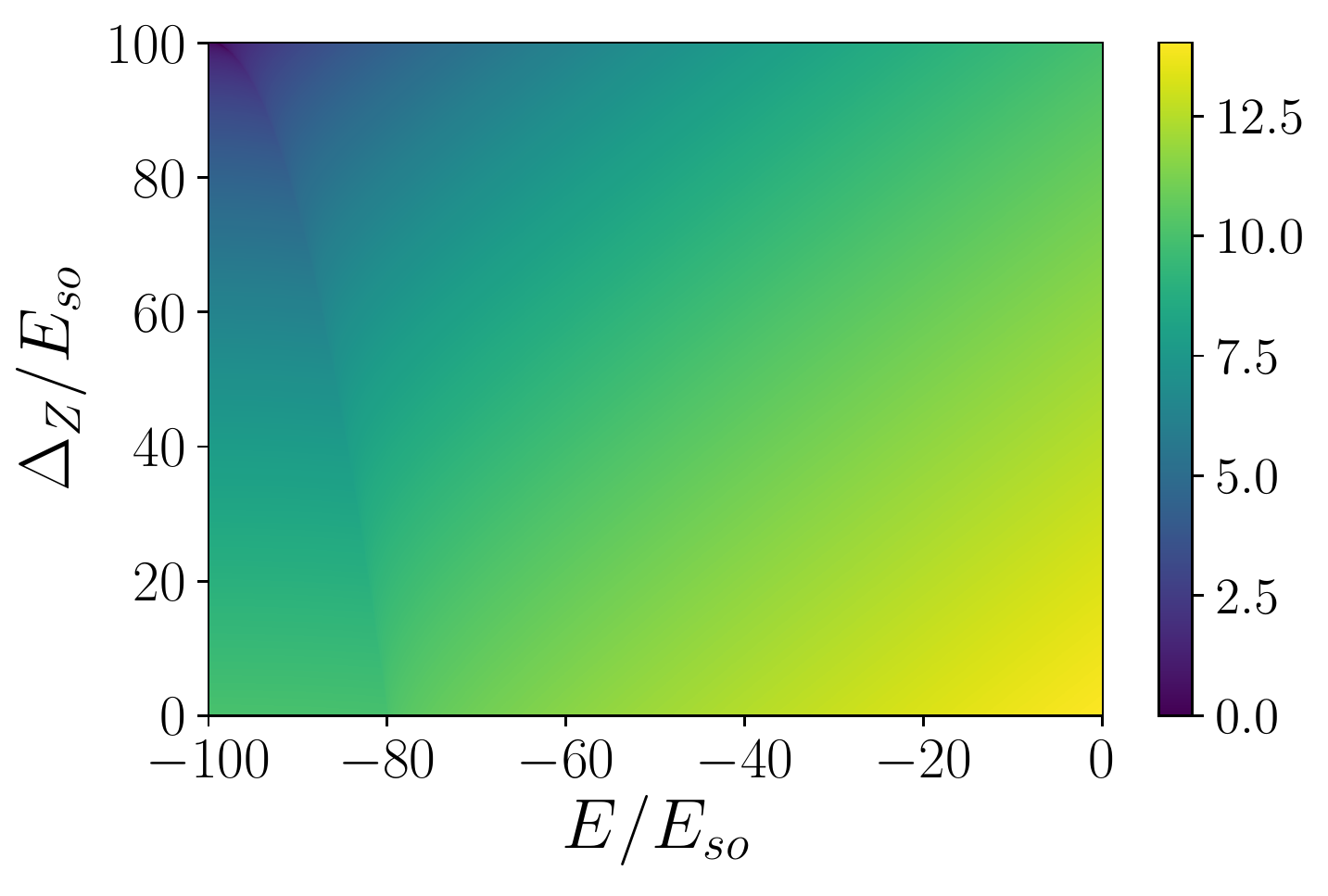}}\\
	\subfloat[Imaginary part of $k_{-,+}/k_{so}$]{\includegraphics[width = 0.3\linewidth]{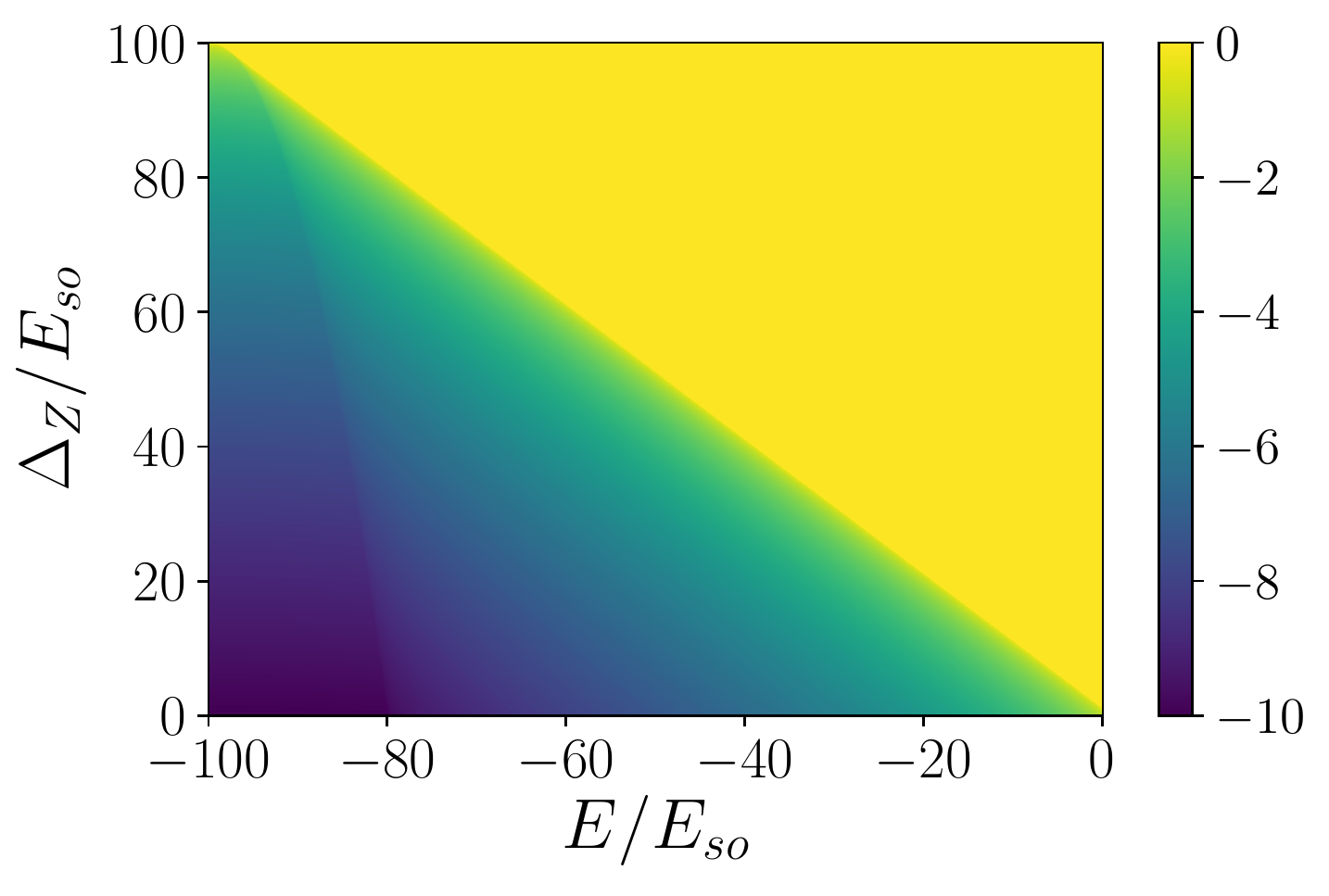}}
	\subfloat[Imaginary part of $k_{+,-}/k_{so}$]{\includegraphics[width = 0.3\linewidth]{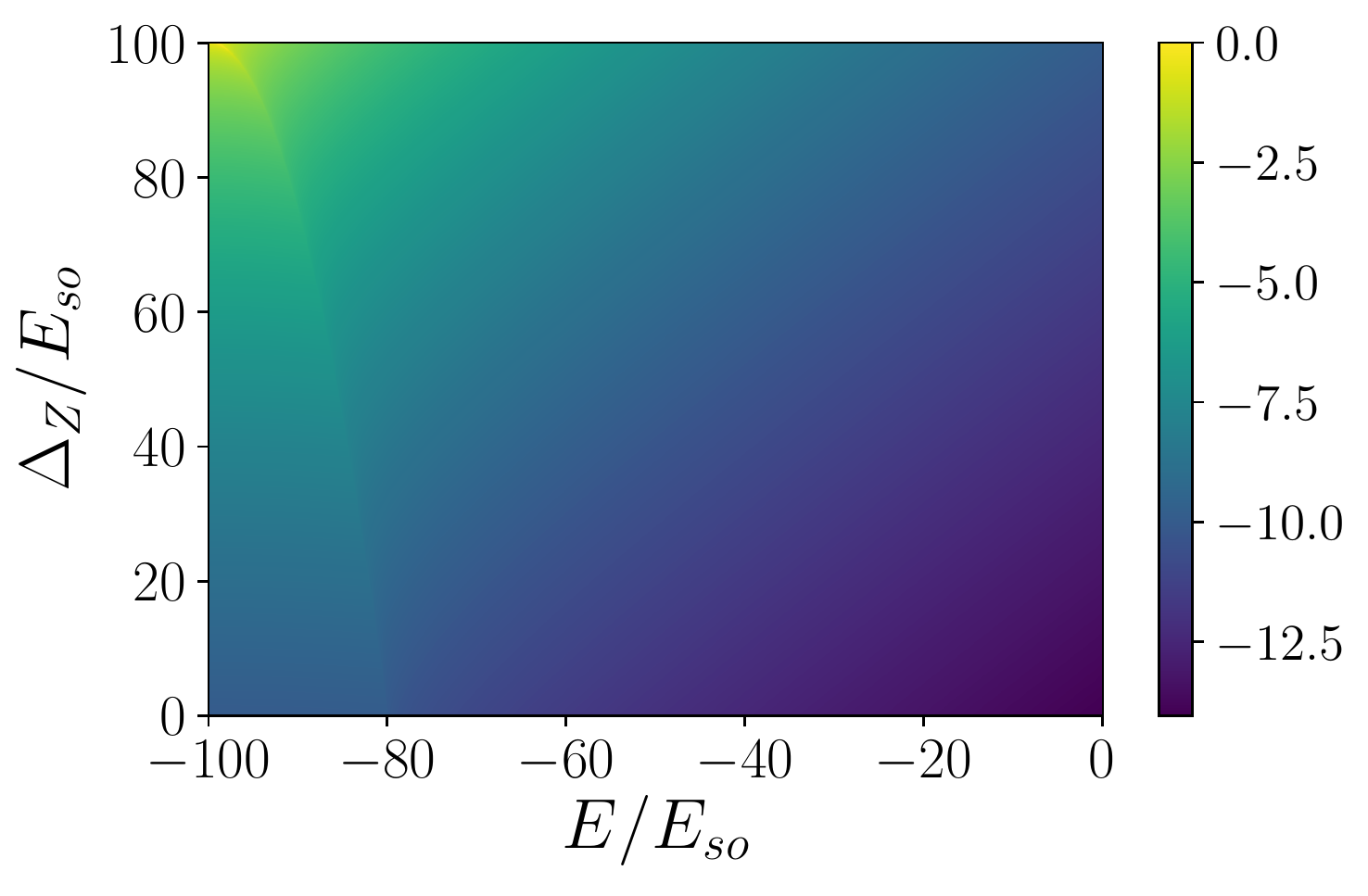}} 
	\caption{Imaginary part of $k_{\pm,\pm}/k_{so}$  as a function of $E/E_{so}$ and $\Delta_Z/E_{so}$. In the first (last) two panels, the imaginary part is always positive (negative). }
	\label{fig:im_part}
\end{figure}

In this section, we provide further details about the solution of the eigenvalue problem set by the Schroedinger equation associated with the Hamiltoniand density 
\begin{equation}
\mathcal{H}(x)=-\left(\frac{\partial_x^2}{2m}+\mu\right)\sigma_0+\Delta_Z(x)\sigma_z-i\alpha_R \partial_x \sigma_y,
\end{equation}
corresponding to the model presented in the main text. Here $\Delta_Z = \Delta_Z \text{sign}(x)$. A particular solution to this equation is provided by the following ansatz
\begin{equation}
\psi(x)=\left(\begin{matrix}
v_{\uparrow}(k)\\v_{\downarrow}(k)
\end{matrix}
\right)e^{i kx},\label{eq:ansatz}
\end{equation}
which assumes that the wavefunction depends on the position $x$ only in this complex exponential. Here, $v_{\uparrow/\downarrow}(k)$ are the component of a two-dimensional spinor in momentum space. The parameter $k$ is a complex number, whose imaginary part is positive (negative) for $x>0$ ($x<0$), in order to find a normalizable solution of Schroedinger equation localized at the boundary $x=0$. With this ansatz, our initial eigenvalues problem is converted into the following
\begin{equation}
\label{eq:Schroedinger2}
\left(\begin{matrix}
\frac{k^2+k_{so}^2}{2m} \pm \Delta _Z & -i \alpha_R k\\i \alpha_R k& \frac{k^2+k_{so}^2}{2m}\mp \Delta_Z
\end{matrix}
\right)\left(\begin{matrix}
v^{(\pm)}_{\uparrow}(k)\\v^{(\pm)}_{\downarrow}(k)
\end{matrix}
\right)=E(k)\left(\begin{matrix}
v^{(\pm)}_{\uparrow}(k)\\v^{(\pm)}_{\downarrow}(k)
\end{matrix}
\right),
\end{equation}
where the signs $\pm$ correspond to $k$ with positive/negative imaginary part. In the above equation, we also fixed the chemical potential to $\mu=-\frac{k_{so}^2}{2m}$. There are four values of $k$ that correspond to a certain energy $E$:
\begin{equation}
k_{\pm,\pm}=\pm k_{so}\sqrt{1+\frac{E}{E_{so}}\pm \sqrt{ \frac{4E}{E_{so}}+\left(\frac{\Delta_Z}{E_{so}}\right)^2}}.
\label{eq:exponent}
\end{equation}
In order to determine the sign of the imaginary part of the above expressions, we investigate the sign of  the imaginary part of $k_{\pm,\pm}/k_{so}$ as a function of $E/E_{so}$ and $\Delta_Z/E_{so}$,  see also Fig. \ref{fig:im_part}. We find that the imaginary part of $k_{+,+}$  and $k_{-,-}$ ($k_{-,+}$and $k_{+,-}$) is always positive (negative). For this reason, $k_{+,+}$  and $k_{-,-}$($k_{-,+}$and $k_{+,-}$) are the correct values for $k$ in the region $x>0$ ($x<0$). It is useful to switch to the following notation:
\begin{align}
k_1\equiv k_{+,+},\hspace{5mm}k_2\equiv k_{-,-},\hspace{5mm}
k_3\equiv k_{+,-},\hspace{5mm}k_4\equiv k_{-,+}.
\end{align}

The corresponding eigenvectors are given by
\begin{subequations}
	\begin{align}
	v_1=\left(\begin{matrix}
	v_{1\uparrow}\\v_{1\downarrow}
	\end{matrix}
	\right)&=\left(\begin{matrix}
	\frac{k_1}{k_{so}}-\sqrt{\left(\frac{k_1}{k_{so}}\right)^2+\left(\frac{\Delta_Z}{2E _{so}}\right)^2}\\\frac{\Delta_Z}{2E_{so}}
	\end{matrix}
	\right),\hspace{5mm}v_3=\left(\begin{matrix}
	v_{3\uparrow}\\v_{3\downarrow}
	\end{matrix}
	\right)=\left(\begin{matrix}
	-v_{1\uparrow}\\v_{1\downarrow}
	\end{matrix}
	\right),\\
	v_2=\left(\begin{matrix}
	v_{2\uparrow}\\v_{2\downarrow}
	\end{matrix}
	\right)&=-\left(\begin{matrix}
	\frac{k_2}{k_{so}}+\sqrt{\left(\frac{k_2}{k_{so}}\right)^2+\left(\frac{\Delta_Z}{2E _{so}}\right)^2}\\\frac{\Delta_Z}{2E_{so}}
	\end{matrix}
	\right),\hspace{4mm}	v_4=\left(\begin{matrix}
	v_{4\uparrow}\\v_{4\downarrow}
	\end{matrix}
	\right)=\left(\begin{matrix}
	-v_{2\uparrow}\\v_{2\downarrow}
	\end{matrix}
	\right).
	\end{align}
	\label{eq:vector}
\end{subequations}
In terms of the particular solution provided by the ansatz in Eq.~\eqref{eq:ansatz}, the general expression for the bound state wavefunction localized at $x=0$ is
\begin{equation}
\psi(x)=\Bigg\{\begin{matrix}
a_1v_1e^{i k_1x}+a_2v_2e^{i k_2x}, \hspace{15mm}x>0\\a_3v_3e^{i k_3x}+a_4v_4e^{i k_4x}, \hspace{15mm}x<0
\end{matrix}\hspace{4mm}.
\end{equation} 
The coefficients $a_1,a_2,a_3,$ and $a_4$ are complex numbers that should be determined by matching the wavefunction at the interface. For a system of second-order partial differential equations,  one has to impose the continuity of wavefunction and its first derivative at $x=0$ at the same time
\begin{align}
\psi(0^+)&=\psi(0^-),\\
\partial_x\psi(x)\Big|_{x=0^+}&=\partial_x\psi(x)\Big|_{x=0^-}.
\end{align}
These boundary conditions can be expressed in terms of a systems of linear equations in the coefficients $a_1,a_2,a_3,$ and $a_4$ as
\begin{equation}
M a=\mathbf{0}_4,
\end{equation}
where $\mathbf{0}_4=(0,0,0,0)^T$, $a=(a_1,a_3,a_2,a_4)^T$ and $M$ is a $4 \times 4$ matrix given by
\begin{equation}
\left(\begin{matrix}
v_{1\uparrow} & v_{2\uparrow} & -v_{3\uparrow} & -v_{4\uparrow}\\v_{1\downarrow} & v_{2\downarrow} & -v_{3\downarrow} & -v_{4\downarrow}\\k_1v_{1\uparrow} & k_2v_{2\uparrow} & -k_3v_{3\uparrow} & -k_4v_{4\uparrow}\\k_1v_{1\downarrow} & k_2v_{2\downarrow} & -k_3v_{3\downarrow} & -k_4v_{4\downarrow}
\end{matrix}\right).
\end{equation}
We note that the only unknown variable of our problem appearing in $M$ is the energy $E$. In order to find a solution to this linear system, one has to impose the vanishing of the determinant of $M$. Indeed, it can be shown that the solution for the above equation exists only for $\frac{\Delta_Z}{E_{so}}<4$, meaning that above this energy threshold no bound state can exist. In total, there are $5$ solutions valid for $\frac{\Delta_Z}{E_{so}}<4$, given by
\begin{align}
\frac{E_1}{E_{so}}&=-\frac{3 \left(\frac{\Delta_Z}{E_{so}}\right)^4+16 \left(\frac{\Delta_Z}{E_{so}}\right)^2-\sqrt{5 \left(\frac{\Delta_Z}{E_{so}}\right)^2-16} \left(\frac{\Delta_Z}{E_{so}}\right)^3-32}{8 \left(\left(\frac{\Delta_Z}{E_{so}}\right)^2+4\right)},\\
\frac{E_2}{E_{so}}&=-\frac{3 \left(\frac{\Delta_Z}{E_{so}}\right)^4+16 \left(\frac{\Delta_Z}{E_{so}}\right)^2+\sqrt{5 \left(\frac{\Delta_Z}{E_{so}}\right)^2-16} \left(\frac{\Delta_Z}{E_{so}}\right)^3-32}{8 \left(\left(\frac{\Delta_Z}{E_{so}}\right)^2+4\right)},\\
\frac{E_3}{E_{so}}&=-\frac{2^{2/3} \left(3 \left(\frac{\Delta_Z}{E_{so}}\right)^2+2\right)}{3 \sqrt[3]{27 \left(\frac{\Delta_Z}{E_{so}}\right)^4+72 \left(\frac{\Delta_Z}{E_{so}}\right)^2+3 \sqrt{81 \left(\frac{\Delta_Z}{E_{so}}\right)^8+48 \left(\frac{\Delta_Z}{E_{so}}\right)^6}+32}}\nonumber+\\&-\frac{1}{12} \sqrt[3]{54 \left(\frac{\Delta_Z}{E_{so}}\right)^4+144 \left(\frac{\Delta_Z}{E_{so}}\right)^2+6 \sqrt{81 \left(\frac{\Delta_Z}{E_{so}}\right)^8+48 \left(\frac{\Delta_Z}{E_{so}}\right)^6}+64}+\frac{2}{3},\\
\frac{E_4}{E_{so}}&=\frac{i \left(\sqrt{3}-i\right) \left(3 \left(\frac{\Delta_Z}{E_{so}}\right)^2+2\right)}{3 \sqrt[3]{54 \left(\frac{\Delta_Z}{E_{so}}\right)^4+144 \left(\frac{\Delta_Z}{E_{so}}\right)^2+6 \sqrt{81 \left(\frac{\Delta_Z}{E_{so}}\right)^8+48 \left(\frac{\Delta_Z}{E_{so}}\right)^6}+64}}\nonumber+\\&-\frac{1}{24} i \left(\sqrt{3}+i\right) \sqrt[3]{54 \left(\frac{\Delta_Z}{E_{so}}\right)^4+144 \left(\frac{\Delta_Z}{E_{so}}\right)^2+6 \sqrt{81 \left(\frac{\Delta_Z}{E_{so}}\right)^8+48
		\left(\frac{\Delta_Z}{E_{so}}\right)^6}+64}+\frac{2}{3},\\
\frac{E_5}{E_{so}}&=-\frac{i \left(\sqrt{3}+i\right) \left(3 \left(\frac{\Delta_Z}{E_{so}}\right)^2+2\right)}{3 \sqrt[3]{54 \left(\frac{\Delta_Z}{E_{so}}\right)^4+144 \left(\frac{\Delta_Z}{E_{so}}\right)^2+6 \sqrt{81 \left(\frac{\Delta_Z}{E_{so}}\right)^8+48 \left(\frac{\Delta_Z}{E_{so}}\right)^6}+64}}\nonumber+\\&+\frac{1}{24} \left(1+i \sqrt{3}\right) \sqrt[3]{54 \left(\frac{\Delta_Z}{E_{so}}\right)^4+144 \left(\frac{\Delta_Z}{E_{so}}\right)^2+6 \sqrt{81 \left(\frac{\Delta_Z}{E_{so}}\right)^8+48
		\left(\frac{\Delta_Z}{E_{so}}\right)^6}+64}+\frac{2}{3}.
\end{align}
\begin{figure}[h]
	\subfloat[Real and imaginary part of $E_1 / E_{so}$]{\includegraphics[width = 0.33\linewidth]{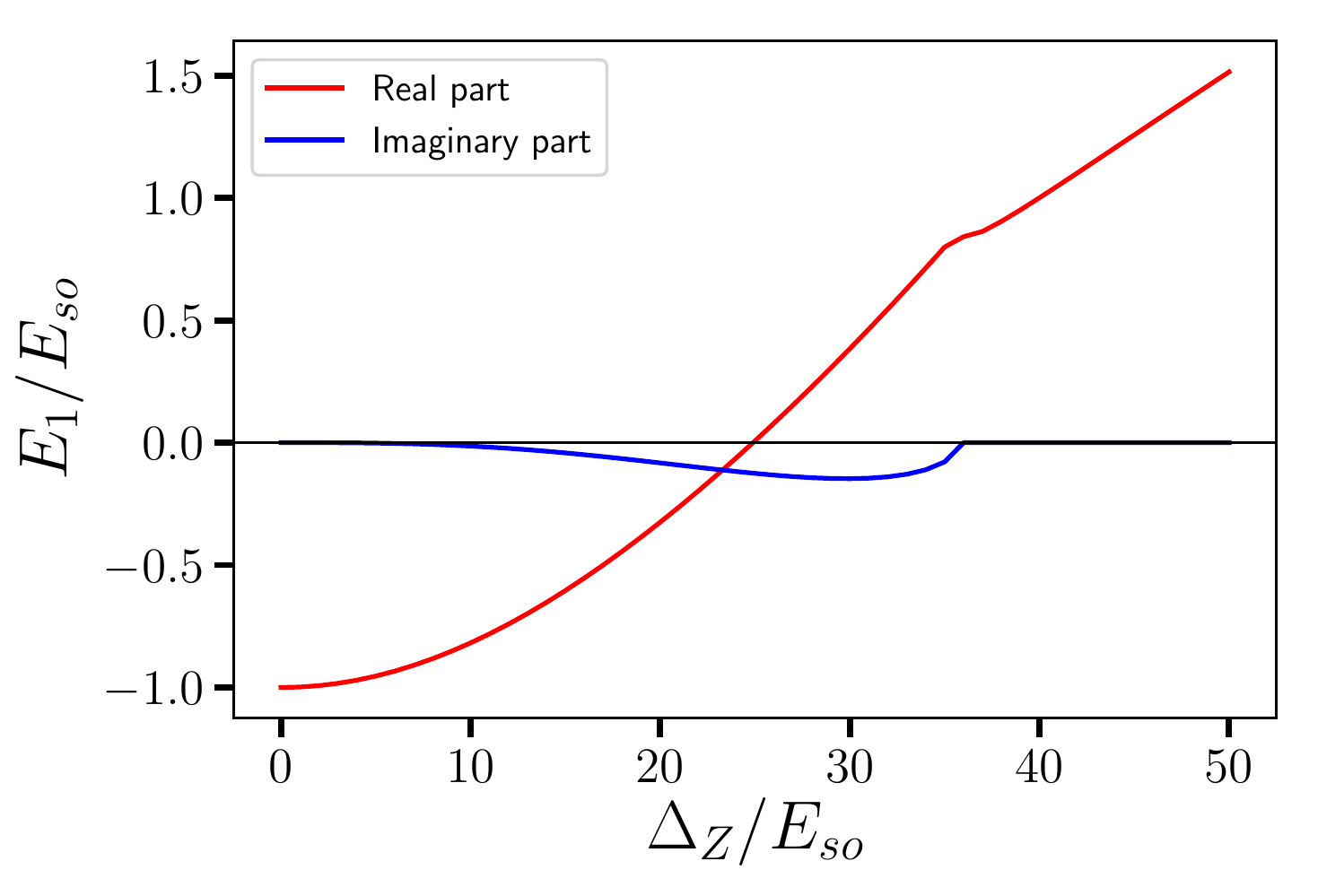}} 
	\subfloat[Real and imaginary part of $E_2 / E_{so}$]{\includegraphics[width = 0.33\linewidth]{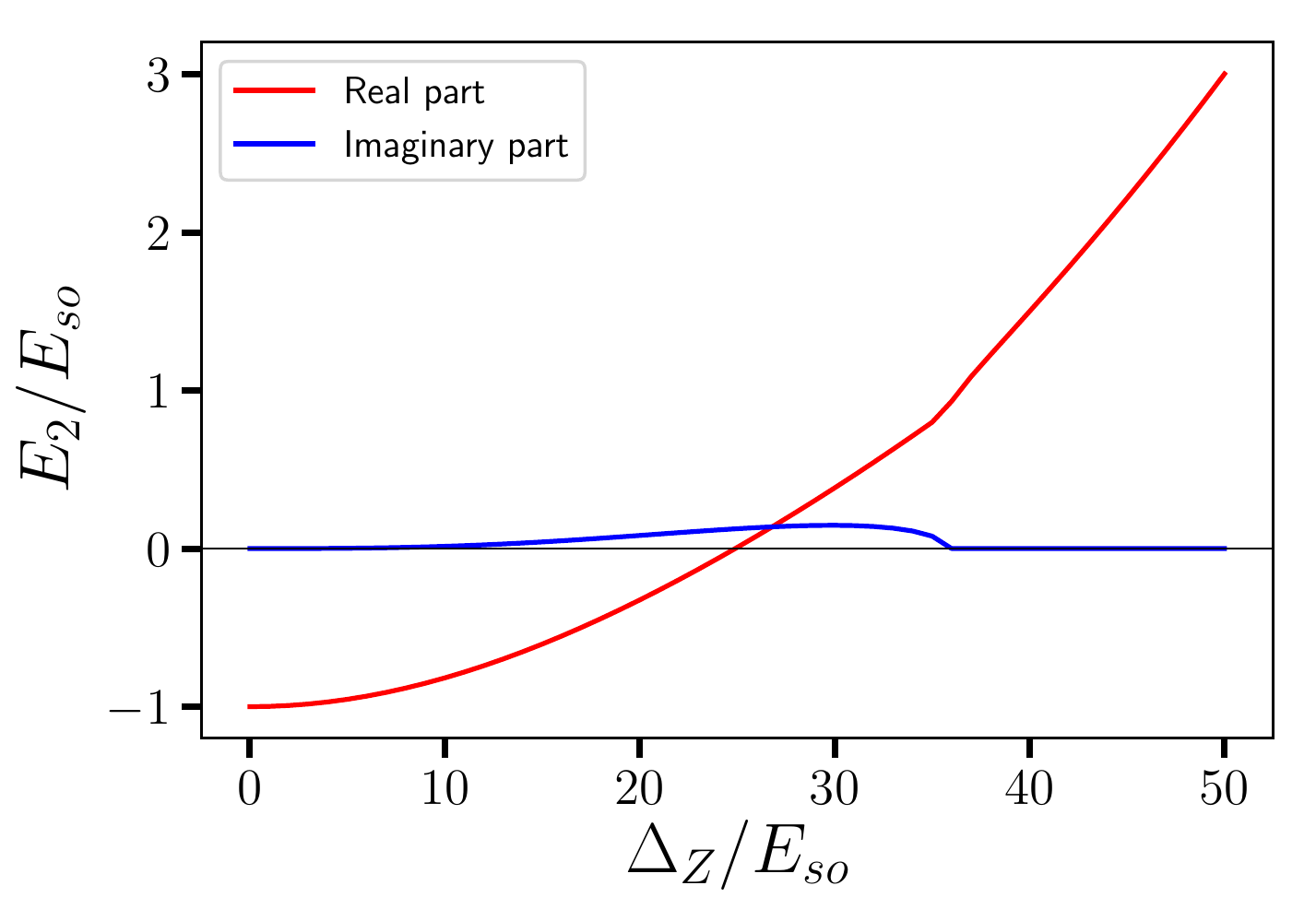}}
	\subfloat[Real and imaginary part of $E_3 / E_{so}$]{\includegraphics[width = 0.33\linewidth]{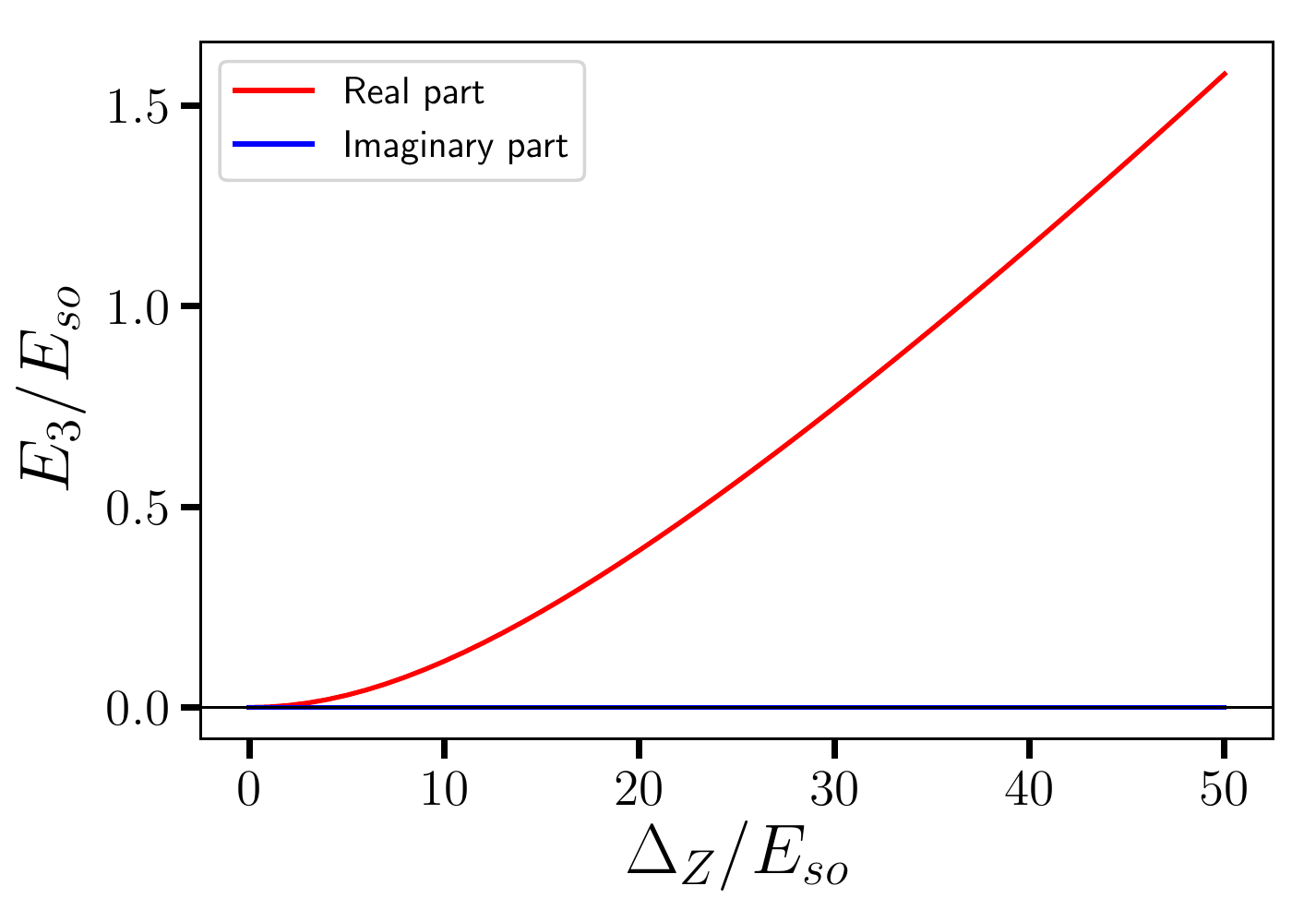}}\\
	\subfloat[Real and imaginary part of $E_4 / E_{so}$]{\includegraphics[width = 0.33\linewidth]{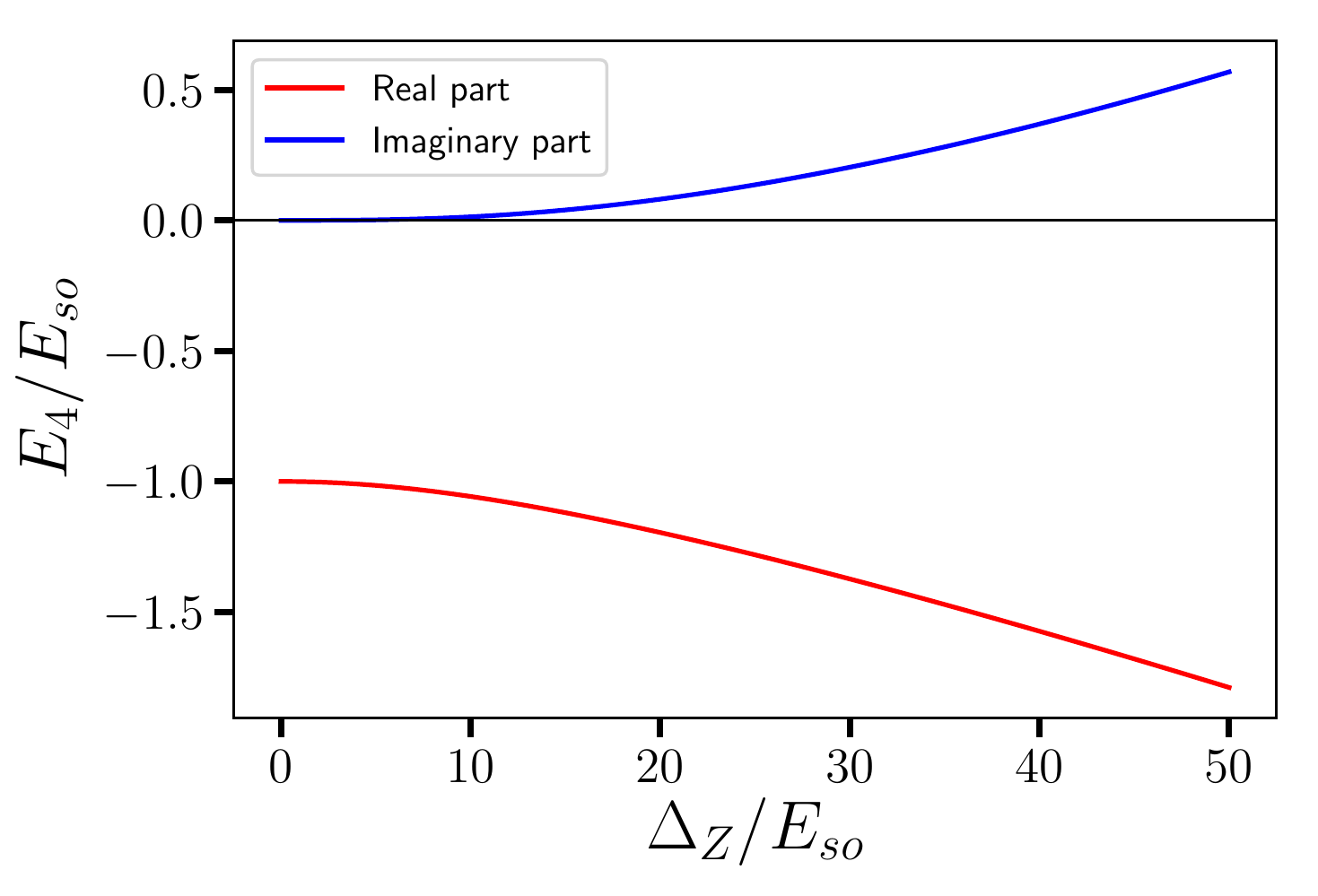}} 
	\subfloat[Real and imaginary part of $E_5 / E_{so}$]{\includegraphics[width = 0.33\linewidth]{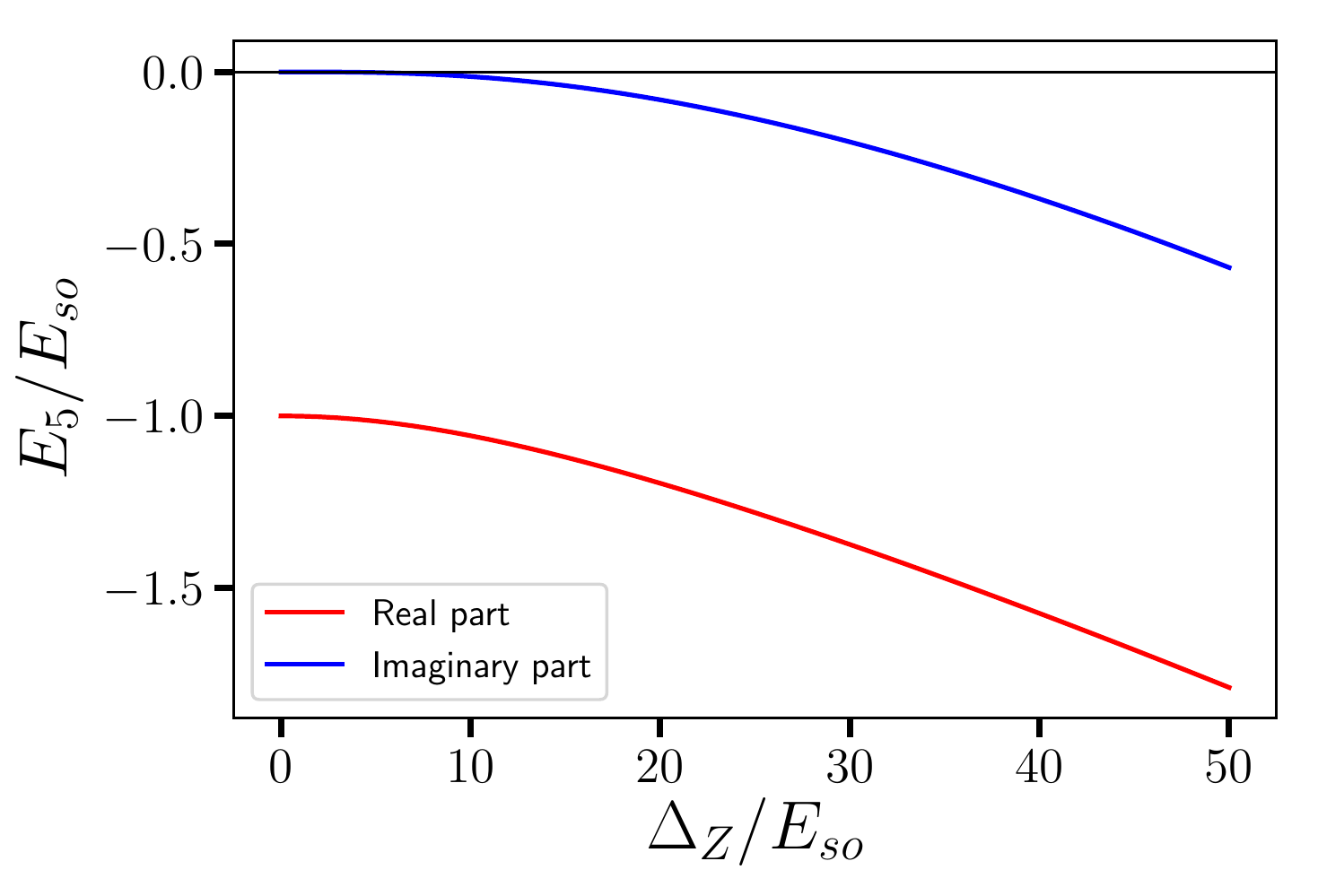}} 
	\caption{Real and imaginary part of $E / E_{so}$ as a function of $\Delta_Z / E_{so}$. Only $E_3$ is  always negative. }
	\label{fig:energy}
\end{figure}
In order to understand which solution is the correct one, we plot in Fig. \ref{fig:energy} the real and imaginary part of each energy as a function of $\Delta_Z / E_{so}$ (which is the only free variable). Clearly, the only physically meaningful solution is $E_3$ since it is the only one which is always real. Therefore, we identify the bound state energy as $E_{BS}\equiv E_3$, thus confirming the result reported in the main text in Eq.~(5).\\ 
Having found the value of $E_{BS}$, it is possible to obtain the value of three coefficients in terms of $k_j$ and $v_j$:
\begin{align}
&\tilde{a}_1=\frac{a_1}{a_4}=\frac{(k_2 v_{2\uparrow} v_{3\uparrow} v_{4\downarrow}-k_2 v_{2\uparrow} v_{3\downarrow} v_{4\uparrow}+k_2 v_{3\uparrow} (v_{2\downarrow} v_{4\uparrow}-v_{2\uparrow} v_{4\downarrow})-k_4 v_{2\downarrow} v_{3\uparrow}
	v_{4\uparrow}+k_4 v_{2\uparrow} v_{3\downarrow} v_{4\uparrow})}{k_1 v_{1\uparrow} (v_{2\uparrow} v_{3\downarrow}-v_{2\downarrow} v_{3\uparrow})+k_2 v_{2\uparrow} (v_{1\downarrow} v_{3\uparrow}-v_{1\uparrow}
	v_{3\downarrow})+k_2 v_{3\uparrow} (v_{1\uparrow} v_{2\downarrow}-v_{1\downarrow} v_{2\uparrow})},\label{eq:coeff1}\\&\hspace{0mm}\tilde{a}_2=\frac{a_2}{a_4}=\frac{ (-k_1 v_{1\uparrow} v_{3\uparrow} v_{4\downarrow}+k_1 v_{1\uparrow} v_{3\downarrow} v_{4\uparrow}-k_2 v_{1\downarrow} v_{3\uparrow} v_{4\uparrow}+k_2 v_{1\uparrow} v_{3\uparrow} v_{4\downarrow}+k_4
	v_{1\downarrow} v_{3\uparrow} v_{4\uparrow}-k_4 v_{1\uparrow} v_{3\downarrow} v_{4\uparrow})}{k_1 v_{1\uparrow} (v_{2\uparrow} v_{3\downarrow}-v_{2\downarrow} v_{3\uparrow})+k_2 v_{2\uparrow} (v_{1\downarrow}
	v_{3\uparrow}-v_{1\uparrow} v_{3\downarrow})+k_2 v_{3\uparrow} (v_{1\uparrow} v_{2\downarrow}-v_{1\downarrow} v_{2\uparrow})},\label{eq:coeff2}\\
&\tilde{a}_3=\frac{a_3}{a_4}=\frac{(k_1 v_{1\uparrow} (v_{2\downarrow} v_{4\uparrow}-v_{2\uparrow} v_{4\downarrow})-k_2 v_{1\downarrow} v_{2\uparrow} v_{4\uparrow}+k_2 v_{1\uparrow} v_{2\uparrow} v_{4\downarrow}+k_4 v_{1\downarrow} v_{2\uparrow}
	v_{4\uparrow}-k_4 v_{1\uparrow} v_{2\downarrow} v_{4\uparrow})}{k_1 v_{1\uparrow} (v_{2\uparrow} v_{3\downarrow}-v_{2\downarrow} v_{3\uparrow})+k_2 v_{2\uparrow} (v_{1\downarrow} v_{3\uparrow}-v_{1\uparrow}
	v_{3\downarrow})+k_2 v_{3\uparrow} (v_{1\uparrow} v_{2\downarrow}-v_{1\downarrow} v_{2\uparrow})},\label{eq:coeff3}\\&a_4=\sqrt{\frac{\left|\tilde{a}_1\right|^2
		\left|v_1\right|^2}{2 {\mathfrak Im}k_1}+
	\frac{\left|\tilde{a}_2\right|^2\left|v_2\right|^2}{2 {\mathfrak Im}k_2}-\frac{\left|\tilde{a}_3\right|^2
		\left|v_3\right|^2}{2 {\mathfrak Im}k_3}-
	\frac{\left|v_4\right|^2}{2 {\mathfrak Im}k_4}},
\end{align}
where the last expression for coefficient $a_4$ is obtained by imposing the normalization condition on $\psi(x)$.

\section{\label{secSm:discreteModel}Discretized model}
\subsection{Nanowire}
In the discretized version of our model, the creation (annihilation) operators
$\psi^{\dag}_{\sigma n }$ ($\psi_{ \sigma n }$) of an
electron with spin component $\sigma$ along the $z$-axis are defined at the discrete coordinate site $m$.
%but we allow for different hopping amplitudes. 
The Hamiltonian describing the Rashba nanowire corresponds
to
\begin{align}
H_{0} = \sum\limits_{n}\Big\lbrace \Big[
- & t_x \left(
\psi^{\dag}_{\uparrow (n+1)} \psi_{\uparrow n} +
\psi^{\dag}_{\downarrow(n+1)} \psi_{\downarrow n}
\right) - \tilde{\alpha}_R \left(
\psi^{\dag}_{\uparrow (n+1) } \psi_{\downarrow n} -
\psi^{\dag}_{\uparrow n} \psi_{ \downarrow (n+1) }
\right)+ \hc
\Big]
+ & \sum\limits_{\sigma}
\left( 2 t_x - \mu \right)
\psi^{\dag}_{\sigma n}
\psi_{\tau \sigma n}
\Big\rbrace
\;.
\end{align}
Here, $t_x = \hbar^2 / (2 m a^2)$, where $a$ is the lattice constant and the spin-flip hopping amplitude $\tilde{\alpha}_R$ is related to the corresponding SOI strengths of the continuum model via
$\alpha_R / \tilde{\alpha}_R = 2 a$~\cite{Reeg}.  The non-uniform Zeeman term is written as
\begin{align}
H_{\trm{Z}}^{\perp} = \sum\limits_{n} \Delta_Z^{(n)} \left(
\psi^{\dag}_{\uparrow n} \psi_{\uparrow n} -
\psi^{\dag}_{\downarrow n} \psi_{\downarrow n}\right)+\hc
\end{align}
The discretized spatial dependence of Zeeman energy is chosen to be 
\begin{equation}
\Delta_Z^{(n)} =g \mu_B B \tanh\left(\frac{n a}{\delta}\right),
\end{equation} 
in order to mimic a smooth transition between the two values of magnetic field parametrized by $\delta$, the width of the domain wall.

\subsection{Two-dimensional layer}
In the discretized version of our two-dimensional model, the creation (annihilation) operators
$\psi^{\dag}_{\sigma m n}$ ($\psi_{\sigma m n}$) of an
electron with spin component $\sigma$ along the $z$-axis are defined at discrete coordinate sites $n$ and $m$. For
simplicity, we assume that the lattice constant $a$ is the same in the $x$ and $y$ directions.
The Hamiltonian describing the 2DEG is given by
\begin{align}
H_{0} = \sum\limits_{mn} \Big\lbrace \Big[
- & t_x \left(
\psi^{\dag}_{\uparrow m (n+1)} \psi_{\uparrow m n} +
\psi^{\dag}_{\downarrow m (n+1)} \psi_{\downarrow m n}
\right) -
t_y
\left( \psi^{\dag}_{\uparrow (m+1) n} \psi_{\uparrow m n} +
\psi^{\dag}_{\downarrow (m+1) n} \psi_{\downarrow m n}
\right)
\notag \\
- & \tilde{\alpha}_x \left(
\psi^{\dag}_{\uparrow (m+1) n} \psi_{\downarrow m n} -
\psi^{\dag}_{\uparrow m n} \psi_{\downarrow (m+1) n}
\right)
+ \tilde{\alpha}_y i \left(
\psi^{\dag}_{\uparrow (m+1) n} \psi_{\downarrow m n} -
\psi^{\dag}_{\uparrow m n} \psi_{\downarrow (m+1) n}
\right) + \hc
\Big]
\notag \\
+ & \sum\limits_{\sigma}
\left( 2 t_x + 2 t_y - \mu \right)
\psi^{\dag}_{\sigma m n}
\psi_{\tau \sigma m n}
\Big\rbrace
\;.
\label{eq:2dmodel1}
\end{align}
Here, $t_x = \hbar^2 / (2 m_x a^2)$ and $t_y = \hbar^2 / (2 m_y
a^2)$. The spin-flip hopping amplitudes $\tilde{\alpha}_x$ and
$\tilde{\alpha}_y$ are related to the corresponding SOI strengths of the continuum model via
$\alpha_y / \tilde{\alpha}_y = \alpha_x / \tilde{\alpha}_x = 2 a$. The non-uniform Zeeman term is given by
\begin{align}
H_{\trm{Z}}^{\perp} = \sum\limits_{mn} \Delta_Z^{(mn)} \left(
\psi^{\dag}_{\uparrow m n} \psi_{\uparrow m n} -
\psi^{\dag}_{\downarrow m n} \psi_{\downarrow m n}\right)+\hc \label{eq:2dmodel2}
\end{align}

\section{\label{secSm:disorder}Stability of Bound States against Disorder}

In this section, we provide additional information about the stability of the bound states found in the Rashba nanowire in the presence of disorder and external  perturbations. While we focus here on the Rashba nanowire, we note that the same stability is also found numerically for the one-dimensional channels localized along the domain wall in the 2D Rashba system.

\begin{figure}[t]
	\includegraphics[width = 0.75\linewidth]{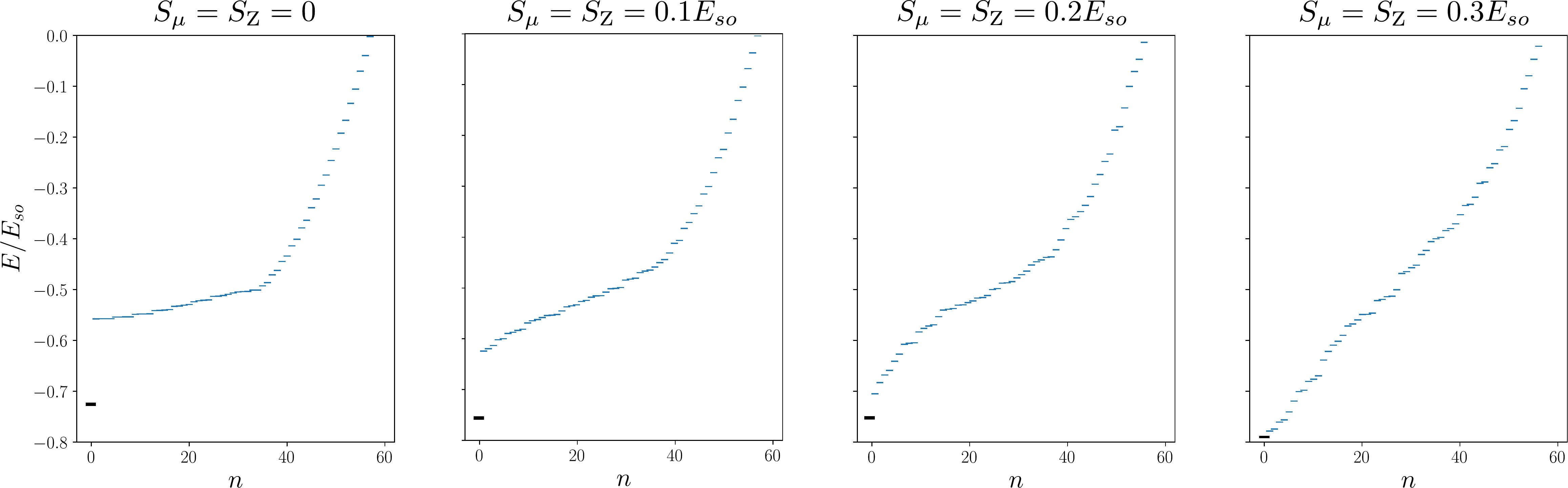}
	\caption{Spectrum of the nanowire in presence of disorder in a chemical potential and Zeeman energy that are fluctuating with standard deviation  $S_{\mu}=S_{\trm{Z}}$  around $\mu=0$ and $\Delta_Z=1.5 E_{so}$, respectively. The bound state is still clearly visible in the spectrum if  the standard deviation  is smaller than the energy separation between the bound state and the bulk modes in the clean limit.}
	\label{fig:disorder}
\end{figure}
\begin{figure}[h]
	\includegraphics[width = .4\linewidth]{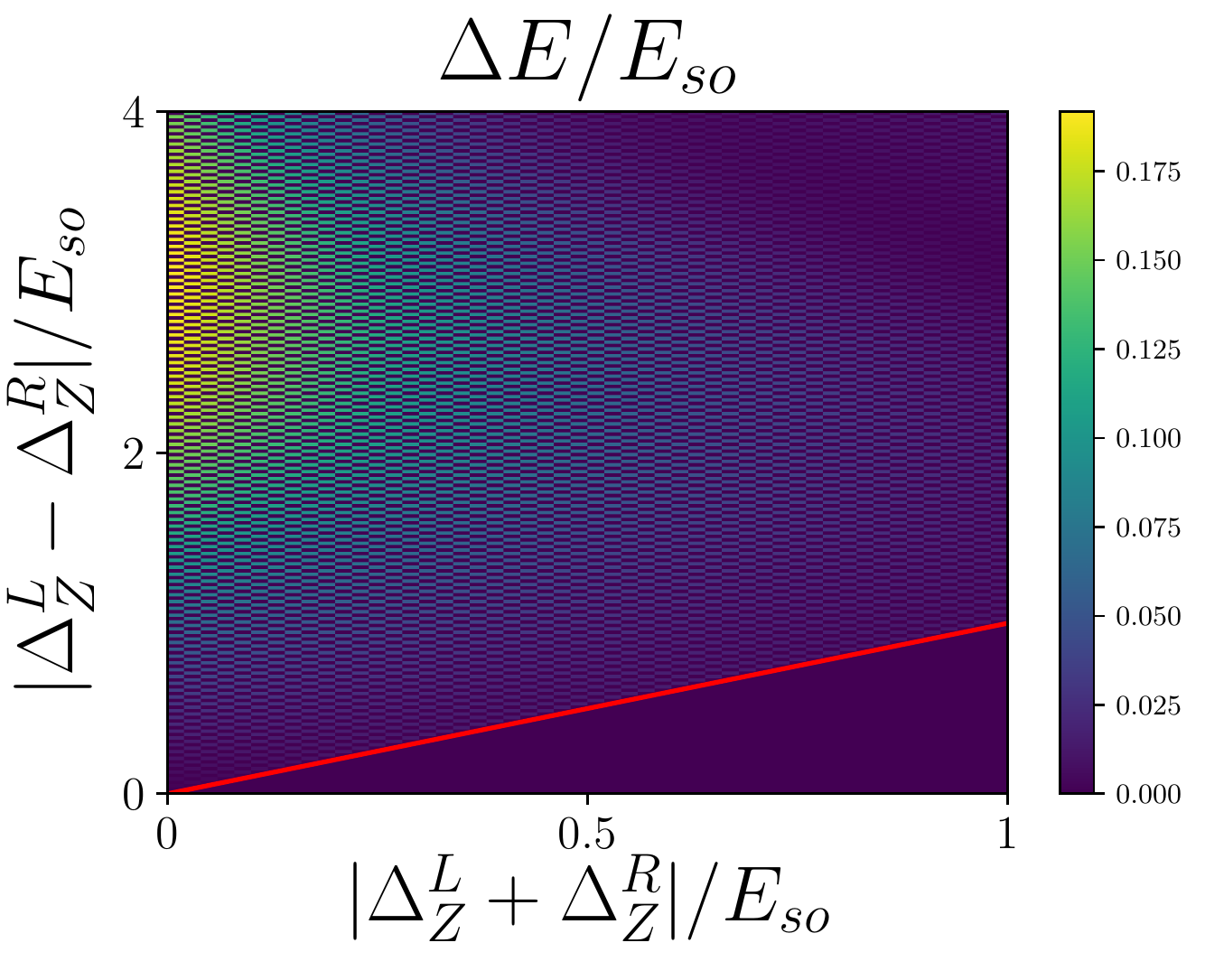}
	\caption{Energy separation $\Delta E / E_{so}$ as a function of absolute values of sum and difference of $\Delta_Z^L$ and $\Delta_Z^R$. The red line corresponds to the case $\Delta_Z^L=0$ or $\Delta_Z^R=0$.}
	\label{fig:B1B2}
\end{figure}
In this part, we show numerically that the bound state is stable against disorder and that it still persists even when the magnitude of the magnetic field is not exactly the same in the two sections. First,  we consider the effect of  disorder such as a fluctuating chemical potential, as well as a disorder in the perpendicular components of the magnetic field. The fluctuating chemical potential and magnetic field have a random amplitude chosen from a uniform distribution with standard deviations $S_{\mu}$ and $S_{\trm{Z}}$, which for simplicity are assumed to be equal. The fluctuating chemical potential mean value is set to zero. In Fig.~\ref{fig:disorder}, we plot the spectrum in the presence of these types of disorder for a mean Zeeman energy $\Delta_Z=1.5 E_{so}$. We observe that the separation between the bound state and bulk states still exists up to a disorder strength of order of $0.2 E_{\trm{so}}$, which is comparable with the energy separation in the clean limit for this value of magnetic field (see the main text). Second, we take into account the possibility of having  different magnitudes of Zeeman energies on the two sides of the domain wall. This assumption is modelled by a non-uniform Zeeman-energy $\Delta_Z(x)=\Delta_Z^L \Theta(-x)+\Delta_Z^R \Theta(x)$. In Fig.~\ref{fig:B1B2}, we plot the energy separation as a function of the absolute values of sum and difference of Zeeman energies $\Delta_Z^L$ and $\Delta_Z^R$. As a result, one can argue that the biggest separation between bulk states and the bound state is achieved for $\Delta_Z^L =  - \Delta_Z^R = 1.5 E_{so}$. The red line in the density plot corresponds to the case where absolute values of sum and difference are equal, meaning that either $\Delta_Z^L$ or $\Delta_Z^R$ vanishes. In the region above this line, the two Zeeman energies have opposite signs. Therefore, one can conclude that the energy separation $\Delta_E$ is still finite even for small deviations in the magnitude of Zeeman energies in the two regions, as long as they have a opposite signs.

\begin{figure}[h]
	\includegraphics[width = \linewidth]{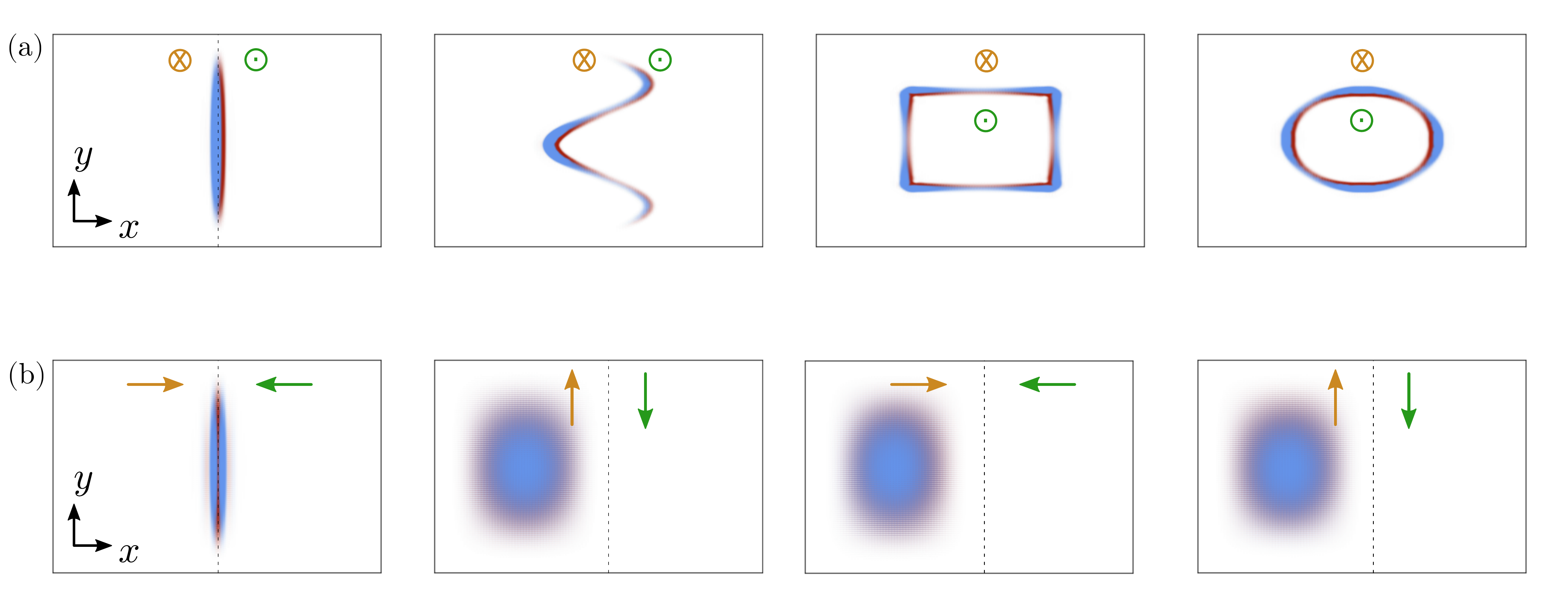}
	\caption{(a) Probability density  for the lowest energy state in the 2D Rashba layer. The magnetic field perpendicular to the 2D layer  changes sign at the boundaries with different shapes: from left to right. These shape are, respectively, a straight line, a sine function, a rectangle, and a circle. (b) Probability density  for the lowest energy state in the 2D Rashba layer for in-plane magnetic fields. In the two leftmost pictures $\alpha_x = 0$, while for the other ones $\alpha_x = \alpha_y\ne0$. The 1D channel is localized around the line $x = 0$ only for the case with $\alpha_x = 0$ for a magnetic field pointing along the $x$ direction.}
	\label{fig:2D}
\end{figure}

\begin{figure}[h]
	\includegraphics[width = .5\linewidth]{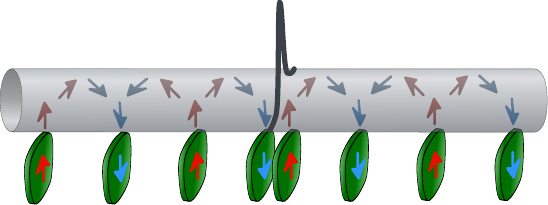}
	\caption{Spatially periodic magnetic texture emerging in a nanowire without Rashba spin-orbit interaction (grey) is created with nanomagnets (green). Red and blue arrows indicate nanomagnets with an opposite magnetization direction. This texture generates an oscillating magnetic field with effectively opposite sign in the two sections of the nanowire, which is equivalent to the configuration presented in the main text. As a result, even in the absence of the Rashba SOI, the bound state (black) can be confined at the interface.}
	\label{fig:nanomagnets}
\end{figure}

\section{Two-dimensional Rashba layer}

Next, we consider different configurations of magnetic domain walls in the case of a two-dimensional Rashba system. In Fig.~\ref{fig:2D}, we plot the probability density  for the lowest energy state found with the model defined in Eqs.~\eqref{eq:2dmodel1} and~\eqref{eq:2dmodel2}. In the two leftmost plots of upper panel, the boundary at which the perpendicular magnetic field changes its sign  assumes two different shapes: a straight line and a sine function. In both cases, the state is localized along this line. Then, in the two rightmost pictures of  Fig.~\ref{fig:2D}(a), the magnetic field changes sign in different regions of the $xy$ plane, which are delimited by closed lines, given, respectively, by a rectangle and by a circle. Even in this case, the probability density of the lowest energy state is localized along the closed boundary where the sign of magnetic field is reversed. In the lower panel, we show the probability density  for the lowest energy state of the Rashba layer for the case of an in-plane magnetic field that changes its sign at the boundary $x=0$. The  SOI vector is pointing along the $y$-axis for the first pair of pictures ($\alpha_x = 0$), while, for the second pair, both components are present ($\alpha_x = \alpha_y\ne0$). The only configuration in which there is a one-dimensional channel localized around the line $x = 0$ is when $\alpha_x = 0$ and the magnetic field points along the $x$ direction. In all the remaining cases, the lowest energy state is a bulk state that extends in one of the two halves of $xy$ plane.

\section{Setup based on the rotating magnetic field }
As noted in the main text, the configuration with a uniform Rashba SOI and a uniform magnetic field applied perpendicular to the SOI vector
is equivalent to a system with a spatially oscillating magnetic texture and no intrinsic Rashba SOI. This opens up an alternative way to generate the domain wall. For example, it can be done by producing a defect in the magnetic texture created by extrinsic nanomagnets \cite{2,3,4}, see Fig.~\ref{fig:nanomagnets}.

\end{document}